\documentclass[fleqn,usenatbib]{mnras}

\usepackage{newtxtext,newtxmath}
\usepackage{graphicx}	
\usepackage{amsmath}	
\usepackage{amssymb}	

\newcommand{\chandra}{\emph{Chandra}}
\newcommand{\xmm}{\emph{XMM-Newton}}

\newcommand{\arcm}{\hbox{$^\prime$}}
\newcommand{\arcs}{\mbox{\arcm\arcm}}
\newcommand{\kmps}{\ensuremath{\mathrm{~km~s^{-1}}}}
\newcommand{\Msol}{\ensuremath{\mathrm{~M_{\odot}}}}
\newcommand{\Zsol}{\ensuremath{\mathrm{~Z_{\odot}}}}

\title[AGN feedback and sloshing in NGC~1550]{Evidence of AGN feedback and sloshing in the X-ray luminous NGC~1550 galaxy group}

\author[K. Kolokythas et al.]{Konstantinos Kolokythas,$^{1}$\thanks{E-mail: k.kolok@nwu.ac.za (KK)}
Ewan O'Sullivan,$^{2}$ Simona Giacintucci,$^{3}$ Diana M. Worrall,$^{4}$\newauthor{Mark Birkinshaw,$^{4}$ Somak Raychaudhury,$^{5,6,7}$ Cathy Horellou,$^{8}$ Huib Intema,$^{9}$  and} \newauthor{Ilani Loubser$^{1}$} 
\\ \\
$^{1}$Centre for Space Research, North-West University, Potchefstroom 2520, South Africa\\
$^{2}$Harvard-Smithsonian Center for Astrophysics, 60 Garden Street, Cambridge, MA 02138, USA\\
$^{3}$Naval Research Laboratory, 4555 Overlook Avenue SW, Code 7213, Washington, DC 20375, USA\\
$^{4}$HH Wills Physics Laboratory, University of Bristol, Tyndall Avenue, Bristol BS8 1TL, UK\\
$^{5}$Inter-University Centre for Astronomy and Astrophysics, Pune 411007, India\\
$^{6}$School of Physics and Astronomy, University of Birmingham, Birmingham B15~2TT, UK\\
$^{7}$Department of Physics, Presidency University, 86/1 College Street, Kolkata 700073, India\\
$^{8}$Chalmers University of Technology, Department of Space, Earth and Environment, Onsala Space Observatory, SE-439 92, Onsala, Sweden\\
$^{9}$Leiden Observatory, Leiden University, Niels Bohrweg 2, 2333 CA Leiden, The Netherlands\\
}

\date{Accepted 2020 May 22. Received 2020 April 21; in original form 2020 February 20}

\pubyear{2020}

\begin{document}
\label{firstpage}
\pagerange{\pageref{firstpage}--\pageref{lastpage}}
\maketitle

\begin{abstract}
We present results from GMRT and \chandra\ observations of the NGC~1550 galaxy group. Although previously thought of as relaxed, we show evidence that gas sloshing and active galactic nucleus (AGN) heating have affected the structure of the system. The 610 and 235~MHz radio images show an asymmetric jet$-$lobe structure with a total size of $\sim$33~kpc, with a sharp kink at the base of the more extended western jet, and bending of the shorter eastern jet as it enters the lobe. The 235$-$610~MHz spectral index map shows that both radio lobes have steep spectral indices ($\alpha_{235}^{610}\geq-1.5$) indicating the presence of an old electron population. The X-ray images reveal an asymmetric structure in the hot gas correlated with the radio structure, as well as potential cavities coincident with the radio lobes, with rims and arms of gas that may have been uplifted by the cavity expansion. The X-ray residual map reveals an arc shaped structure to the east that resembles a sloshing cold front. Radio spectral analysis suggests a radiative age of about  33~Myr for the source, comparable to  the sloshing timescale and dynamical estimates of the age of the lobes. An estimate of the mechanical energy required to inflate the cavities suggests that the AGN of NGC~1550 is capable of balancing radiative losses from the intragroup medium (IGM) and preventing excessive cooling, providing that the AGN jets are efficiently coupled to the IGM gas. In conclusion, we find evidence of sloshing motions from both radio and X-ray structures, suggesting that NGC~1550 was perturbed by a minor merger or infalling galaxy about 33~Myr ago.

\end{abstract}




\section{Introduction}

It is now well known that the central gas in the cores of both galaxy clusters and groups is heated by some mechanism. The evidence for this comes from the central temperatures which are found to be significantly higher than the ones predicted by radiative cooling models \citep{Fabianetal94,PetersonFabian06} for the hot X-ray emitting intragroup/intracluster medium (IGM/ICM) gas that is cooling  on timescales shorter than the Hubble time \citep[e.g.,][]{Sandersonetal06}. A potential source of heating that is capable of compensating for the observed radiative losses (feedback) and regulating gas cooling, is energy injection by the central region of an active galactic nucleus \citep[AGN;][]{McNamaraNulsen07}. However, although several studies have shown that heating from AGN feedback is sufficient to balance cooling \citep[e.g.,][]{Gittietal07,Giacintuccietal11,OSullivanetal11b}, the nature of the AGN feedback heating mechanism and the details of energy transfer from the radio jet to the ambient ICM/IGM are yet to be understood.

The impact on the thermal state of a group/cluster of dynamical or thermodynamic disturbances, such as mergers or outbursts of the central AGN, is imprinted in the complex morphology of the X-ray substructures in the IGM/ICM. Physical processes, such as bulk motions and shocks induced by mergers, generate turbulence in certain areas that expand and heat up the surrounding ICM/IGM gas \citep{Bykov15}. A merging event may also disrupt or trigger a radio AGN outburst. A deeper understanding of the events taking place in the intragroup medium along with the effect on radiative outbursts emanating from the central region is provided by the study of systems using both X-ray and radio data. Such studies have revealed cavities in the X-ray brightness in clusters \citep[e.g.,][]{Birzanetal04,Gittietal07,Fabianetal11,McNamaraNulsen12} and groups \citep[e.g.,][]{OSullivanetal11c,Davidetal11,OSullivanetal17,Schellenbergeretal17,Wenhaoetal19}.

The galaxy$-$group environment, although not as massive or X-ray luminous as galaxy clusters, presents a gas fraction that varies significantly (by a factor of $\sim$2 at any given temperature) within r$_{2500}$ compared to clusters \citep{Randalletal09}, with gas fraction being closely correlated to short central cooling time rather than entropy \citep{Gastaldelloetal07,Sunetal09}.  Hence, due to the shallower gravitational potential and low central density, the effects of heating by a central AGN can be pronounced in the group environment. However, following the hierarchical growth model, galaxy groups and clusters in the local Universe are also constantly involved in an efficient gravitational process that can affect the central gas and transform their galaxies: mergers \citep[e.g.,][]{Toomres72}. Through this process, clusters grow via sequential mergers and accretion of smaller systems --- subclusters, galaxy groups and galaxies. Groups grow mainly through the infall of individual galaxies. As galaxy groups are found to merge at a rate higher than clusters ($\sim$2 orders of magnitude more often; \citealt{Mamon00}) they are an ideal environment and an excellent opportunity to study both gravitational (e.g., tidal interactions, sloshing) and non-gravitational (AGN feedback heating) processes that take place in the centres of groups, where there is more, and faster, galaxy evolution and heating is more important \citep[e.g.,][]{MulchaeyZabludoff98,HashimotoOemler00}.

 During a merger event, `sloshing' is the most common phenomenon seen in \textit{Chandra} observations of both groups \citep{Machaceketal11} and clusters of galaxies (\citealt{MarkevitchVikhlinin07}, e.g., A3560;\citealt{Venturietal13}) . In the case of galaxy groups, gas  sloshing takes place when a galaxy (perturber) passes near the central galaxy in an off-axis interaction \citep{AscasibarMarkevitch06}. In particular, this kind of interaction can offset the gas surrounding the core causing its oscillation around the centre of the dark matter potential, leading to a spiral morphology visible in both the temperature and density structures \citep{AscasibarMarkevitch06}. The characteristic feature of sloshing is a cold front, or contact discontinuity, with such a spiral morphology. Both the temperature and density of the gas jump at the sloshing front, but the pressure profile typically remains continuous. The investigation of cold fronts and especially sloshing effects is of great significance as both have an impact on gas heating, enrichment, blending, and turbulence \citep[e.g., as in NGC~5098;][]{Randalletal09}. However, although several cold fronts have been identified by \textit{Chandra} in clusters and groups, sloshing cold fronts have been revealed in fewer systems (e.g., \citealt{Mazzottaetal01}; \citealt{Dupkeetal07} \citealt{Gastaldelloetal09}, \citealt{Randalletal09}). The mechanism of sloshing has been investigated numerically in several studies \citep[e.g.,][]{AscasibarMarkevitch06,ZuHoneetal11,Roedigeretal11}. 

In this paper we describe a combined radio--X-ray study of the nearby (\textit{z}=0.0124) galaxy group centred on NGC~1550. This group has a mass M$_{500}$=3.2$^{+0.3}_{-0.4}\times10^{13}\Msol$ \citep{Sunetal09} and is one of the brightest groups of galaxies in the X-ray sky \citep{Sunetal03}. NGC~1550 was classified as a fossil group by \citet{Jonesetal03} indicating that the galaxy population is strongly dominated by the central elliptical. However, \citet{Sunetal03} showed that it does not meet the definition of the class, having several galaxies $<$2 mag fainter in $R$-band within 0.5R$_{\rm vir}$. Nonetheless, NGC~1550 has been considered a relaxed system, with no evidence of dynamical interactions. \citet{Dunnetal10} observed the NGC~1550 galaxy with the VLA at 1.4~GHz, finding an asymmetric jet extending roughly east-west. However, the spatial resolution of the data was such that no clear lobes were visible, but only two blobs of bright emission in the jet, one overlapping the nucleus and another one extending to the west.


We investigate the radiative properties and  dynamical state of NGC~1550, presenting results from the Giant Metrewave Radio Telescope (GMRT) at 235/610~MHz and archival X-ray observations from the \chandra\ X-ray observatory. We determine the radio source's radiative age and examine in detail its energetics and the dynamical properties of the X-ray emitting gas. NGC~1550 is part of the Complete Local-Volume Groups Sample (CLoGS). An overall description of CLoGS and the X-ray properties of the high--richness sub-sample are given in \citet{OSullivanetal17}. See also the description of the radio properties in \citet{Kolokythasetal18}. NGC~1550 was included in the analysis of the radio properties of the CLoGS low--richness sub-sample \citep{Kolokythasetal19} and the GMRT 235 and 610~MHz images were presented in that paper. Those images form the basis of the more detailed analysis presented in this work.

We organize the paper as follows. In Section 2 we describe the GMRT observations along with the X-ray and radio data analysis. In Section 3 we present results from the radio and X-ray observations, and provide a radio spectral index map, perform spectral ageing modeling fits, examine the distribution of the X-ray emission, and estimate the effect of the radio jets from the cavity energetics. In Section 4 we discuss the radio source's age along with the processes that have taken place during its development, suggesting the most likely scenario for its current state and morphology. Lastly, in Section 5 we present the conclusions of this study. 

We adopt a redshift of $z=0.0124$ and a distance of 53~Mpc for the source (as in \citealt{OSullivanetal17} and \citealt{Kolokythasetal19}) giving an equivalent angular scale of 0.257 kpc arcsec$^{-1}$.



\begin{table*}
\begin{minipage}{\linewidth}
 \caption{Details of GMRT observations for NGC~1550. The columns give the observation date, frequency, time on source, beam parameters and the rms noise in the resulting image.}
 \centering
 \label{GMRTtable}
\begin{tabular}{|c|c|c|c|c|}
 
 \hline \hline
 Observation date & Frequency &    On source time      &   Beam, P.A.                  & rms \\
 
      &  (MHz)    &   (minutes)    & (Full array $''\times'',{}^{\circ}$) & (mJy beam$^{-1}$) \\
 \hline
 2011 Dec  &    610    &   187 &  $5.66\times3.96$, 89   & 0.04 \\ 
2011 Dec  &    235    &   187   &  $14.28\times10.65$, 89  & 0.45 \\ 

 \hline
 \end{tabular}
\end{minipage}
 \end{table*}

\section{Observations and data reductions}
\subsection{Giant Metrewave Radio Telescope}
\label{sec:gmrt}
NGC~1550 was observed for $\sim$3 hours (on source time) using the GMRT in dual 235/610~MHz frequency mode, in cycle~21, during 2011 December. The data for both frequencies were recorded using the upper side band (USB) correlator  providing an observing bandwidth of 32 MHz at both 610 and 235~MHz. 

Data collection was performed using 512/256 channels, at a spectral resolution of 65.1/130.2 kHz per channel, for the 610/235~MHz bands, respectively. A summary of the observations can be seen in Table~\ref{GMRTtable}. 

The data were processed as described in \citet{Kolokythasetal19}, using the \textsc{spam} pipeline\footnote{For more information  on how to download and run \textsc{spam} see \url{http://www.intema.nl/doku.php?id=huibintemaspam}}, a \textsc{python} based extension to the NRAO Astronomical Image Processing System (\textsc{aips}) package \citep{Intema14}. Only a brief summary of the data analysis procedure is given here. For more details regarding the \textsc{spam} pipeline and the algorithms of the \textsc{spam} package see \citet{Intemaetal09,Intemaetal17}.


The first stage of the \textsc{spam} pipeline converts the raw LTA format data collected from the observations into pre-calibrated visibility data sets (UVFITS format). The second stage converts these visibilities into a final Stokes~I image (in FITS format), via several repeated steps of (self)calibration, flagging, and wide-field imaging. This stage includes direction-dependent calibration, radio frequency interference (RFI) mitigation, imaging and ionospheric modeling that adjusts for the dispersive delay in the ionosphere \citep{Intemaetal17}.

The flux density scale was set from the flux calibrator (3C~147) using the model of \citet{ScaifeHeald12}. We adopt an uncertainty on the flux density measurements of 5\% at 610~MHz and 8\% at 235~MHz to account for residual amplitude calibration errors \citep{Chandra04}.

\subsection{Chandra}
\label{sec:chandra}
NGC~1550 has been observed by \chandra\ four times, twice using the Advanced CCD Imaging Spectrometer (ACIS) in its I configuration, and twice with ACIS-S. A summary of the \chandra\ mission is provided in \citet{Weisskopfetal02}, and Table~\ref{tab:Xray} contains a summary of the observations. All four observations were performed in VFAINT data mode. The data were processed using \textsc{ciao} 4.11 and CALDB 4.8.2. Our reduction followed the approach laid out in the \chandra\ analysis threads\footnote{http://cxc.harvard.edu/ciao/threads/index.html} and \citet{OSullivanetal17}. 

\begin{table}
\caption{\label{tab:Xray}Summary of the \chandra\ observations.}
    \centering
    \begin{tabular}{lccc}
    \hline
    ObsID & Observation date& Instrument & Cleaned/uncleaned \\
          &                 &            & exposure (ks) \\
    \hline
    3186 & 2002 Jan 08 & ACIS-I & 9.7/10.0 \\
    3187 & 2002 Jan 08 & ACIS-I & 9.1/9.7 \\
    5800 & 2005 Oct 22 & ACIS-S & 44.5/44.5 \\
    5801 & 2005 Oct 24 & ACIS-S & 44.5/44.5 \\
    \hline
    \end{tabular}
\end{table}

Very faint mode filtering was applied to all datasets, and periods of high background (flaring) were filtered out using the \textsc{lc\_clean} script. We used the standard set of \chandra\ blank-sky background files to create background spectra and images, normalizing to the 9.5-12~keV count rate of the observations. All observations were reprojected onto a common tangent point, and combined images and exposure maps created using \textsc{reproject\_obs} and \textsc{merge\_obs}. Point sources were identified using the \textsc{wavdetect} task, and removed. For imaging analysis we used the combined 0.5-2~keV images. Spectra and responses were extracted from each observation individually, and combined for front and back illuminated chips separately. Surface brightness modelling was performed in \textsc{ciao sherpa} v4.11 \citep{Freemanetal01}.

Spectral fitting was performed in \textsc{Xspec} v12.10.1 \citep{Arnaud96}. We adopted a hydrogen column of 1.02$\times$10$^{21}$~cm$^{-2}$, drawn from the Leiden-Argentine-Bonn survey \citep{Kalberlaetal05}. We adopted the solar abundance ratios of \citet{GrevesseSauval98}.

\begin{table*} 
\caption{Radio flux densities, radio power and spectral indices of NGC~1550. The columns list the total flux density at 235 and 610~MHz, the 235$-$610~MHz spectral index, the flux density at 1.4~GHz (drawn from the literature), the 235$-$1400~MHz spectral index,  the flux density at 150~MHz (drawn from the literature), the 150$-$235~MHz and the 150$-$610~MHz spectral indices, the radio power at 235 and 610~MHz and the largest projected size of the radio source at 235~MHz.
\label{Sourcetabletotal}}
\begin{center}
\begin{tabular}{ccccccccccc}
\hline 
  S$_{235 \rm{MHz}}$ & S$_{610 \rm{MHz}}$ & $\alpha_{235}^{610}$ & S$_{1.4\rm{GHz}}$ & $\alpha_{235}^{1400}$ &  S$_{150 \rm{MHz}}$ & $\alpha_{150}^{235}$ & $\alpha_{150}^{610}$  & P$_{235 \rm{MHz}}$ & P$_{610\rm{MHz}}$ & Projected size \\ 
  $\pm8\%$ (mJy) & $\pm5\%$ (mJy) &   ($\pm$0.04)   &   (mJy)     &   ($\pm$0.06) &  & ($\pm$0.06) & ($\pm$0.05)  & (10$^{23}$ W Hz$^{-1}$) & (10$^{23}$ W Hz$^{-1}$) & (kpc $\times$ kpc) \\
\hline

223.0   &    62.0  &  -1.34 &  17$\pm2^a$  &  -1.44 &  404$\pm$41$^b$  & -1.38  &  -1.35    & 0.752  & 0.209 & 32.6 $\times$ 15.8\\
\hline
\end{tabular}
\end{center}
$^a$ \citet{Brownetal11}
$^b$ Calculated from the mosaic of TGSS-ADR \citep{Intemaetal17}
\end{table*}

\begin{table*} 
\caption{GMRT radio flux densities, spectral index and projected area of the components in NGC~1550. The columns list the flux density at 235 and 610~MHz, the 235$-$610~MHz spectral index and the projected area of the core, the eastern and western lobes at 235 and 610~MHz respectively. \label{Sourcetablecomp}}
\begin{center}
\begin{tabular}{lccccc}
\hline 
Component & S$_{235 \rm{MHz}}$ & S$_{610 \rm{MHz}}$ & $\alpha_{235}^{610}$ & Projected Area & Projected Area \\ 
 & $\pm8\%$ (mJy) & $\pm5\%$ (mJy) &  $\pm$0.04 &  at 235~MHz (kpc$^2$) &  at 610~MHz (kpc$^2$)\\
\hline
Core      & $\leq4.3$  &  $\leq3.4$  & $-$  & $\leq4$  & $\leq3$ \\
East jet  & 8.3  &  2.6  & $-1.22$  & 9  & 5 \\
West jet  & 17.1 &  5.6  & $-1.17$  & 34 & 24 \\
East lobe & 118.5&  25.9 & $-1.59$  & 135& 58 \\
West lobe & 54.5 &  12.7 & $-1.53$  & 86 & 45 \\
\hline
\end{tabular}
\end{center}
\end{table*}


\begin{table} 
\caption{Integrated radio flux densities for NGC~1550 drawn from the literature. \label{Integratedspix}}
\begin{center}
\begin{tabular}{ccc}
\hline 
$\nu$  & Flux density & Reference \\ 
 (MHz) &  (mJy)       &     \\
\hline
88   & 742$\pm$59$^a$  &  \citet{HurleyWalkeretal17} \\
119   & 523$\pm$42$^a$  &  \citet{HurleyWalkeretal17} \\
155   & 405$\pm$32$^a$  &  \citet{HurleyWalkeretal17} \\
201   & 304$\pm$24$^a$  &  \citet{HurleyWalkeretal17} \\
150   & 404$\pm$41$^b$  &  \citet{Intemaetal17} \\
1400  & 17$\pm$2$^c$  &  \citet{Brownetal11} \\
2380  & 8$\pm$3$^c$ & \citet{DresselCondon78}  \\
\hline
\end{tabular}
\end{center}
$^a$ Calculated from the mosaic of GLEAM survey\\
$^b$ Calculated from the mosaic of TGSS-ADR\\
$^c$ Extracted from NED$^4$
\end{table}

\begin{table} 
\caption{Synage++ spectral aging model fit parameters for power-law, CI and JP models for the integrated spectrum of NGC~1550 shown in Figure~\ref{fig:integratedfit2}. \label{tab:integratedfits}}
\begin{center}
\begin{tabular}{cccc}
\hline 
Model Fits & Break frequency & $\alpha_{\mathrm{inj}}$   & Reduced $\chi^2$\\ 
          &   (MHz)           &                  &     \\ 
\hline
 power-law &     -    &    1.37$_{-0.06}^{+0.01}$      &  0.67       \\
  CI      &  457$_{-394}^{+139}$   & 1.02$_{-0.17}^{+0.06}$  & 0.37  \\
  JP      &   4601$_{-2912}^{+1210}$  &   1.07$_{-0.20}^{+0.03}$  & 0.36  \\
\hline
\multicolumn{4}{l}{Excluding 2.38~GHz}\\
\hline
power-law &       -                 & 1.37$_{-0.06}^{+0.01}$      & 0.76  \\
 CI      &       -                 & 1.35$_{-0.05}^{+0.01}$      & 1.13 \\
 JP      & 3714$_{-2298}^{+375}$ & 1.04$_{-0.21}^{+0.05}$  & 0.35  \\
\hline
\end{tabular}
\end{center}
\end{table}

\section{Results}

\subsection{Radio images and source morphology}
\label{sec:Radio}

\begin{figure*}
\includegraphics[width=0.506\textwidth]{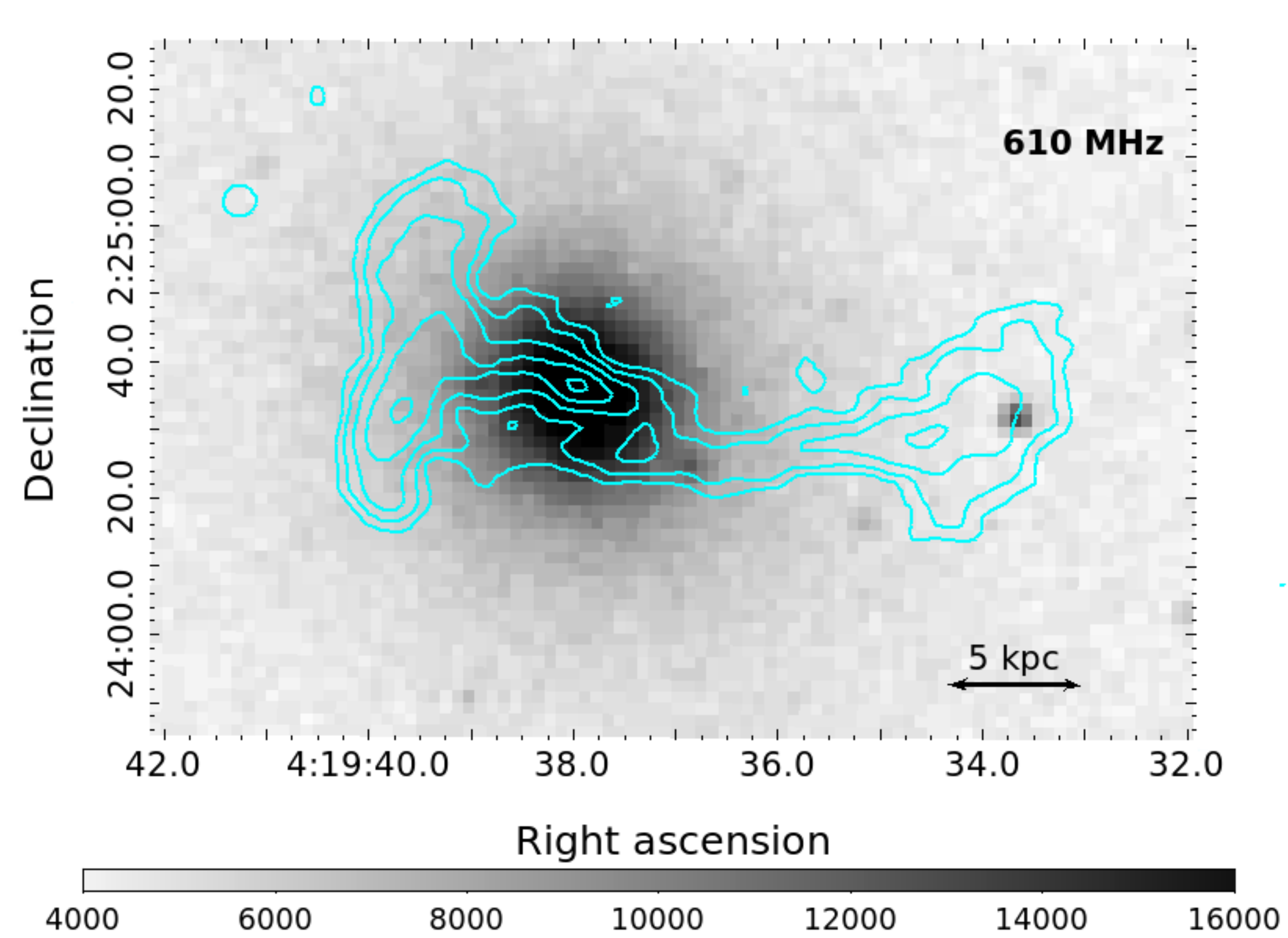}
\includegraphics[width=0.49\textwidth]{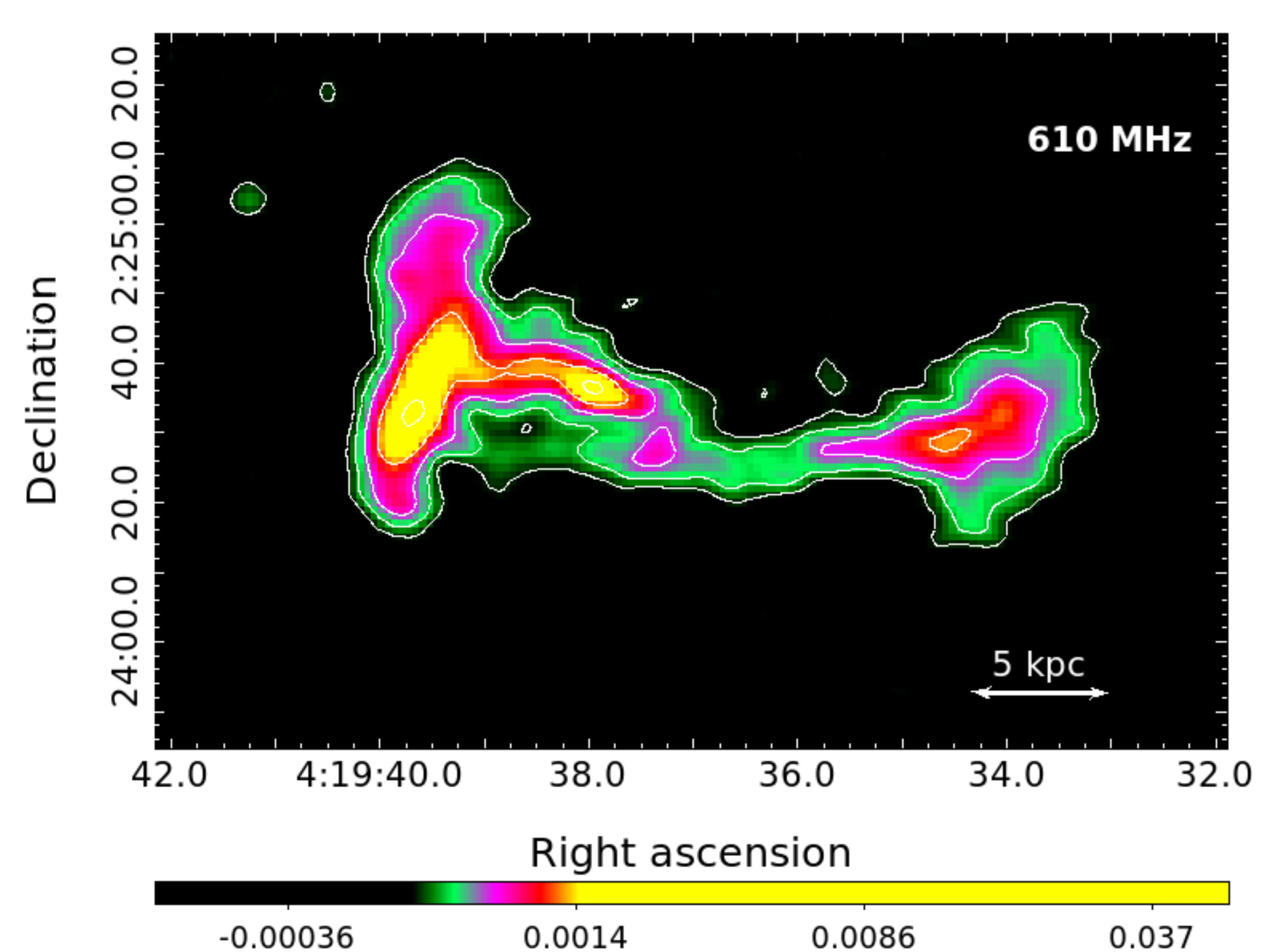}
\caption{\label{fig:radiocont610}\textit{Left:} GMRT 610~MHz radio contours overlaid on a DSS optical image of NGC~1550. \textit{Right:} GMRT 610~MHz radio image and contours of the galaxy. In both panels Right ascension is on the x-axis and Declination is on the y-axis. Contour levels start at 3$\sigma$ and rise in steps of a factor of 2, with the 1$\sigma$ rms noise being $\sim$ 0.04~mJy beam$^{-1}$ at this frequency.}
\end{figure*}

\begin{figure*}
\includegraphics[width=0.5\textwidth]{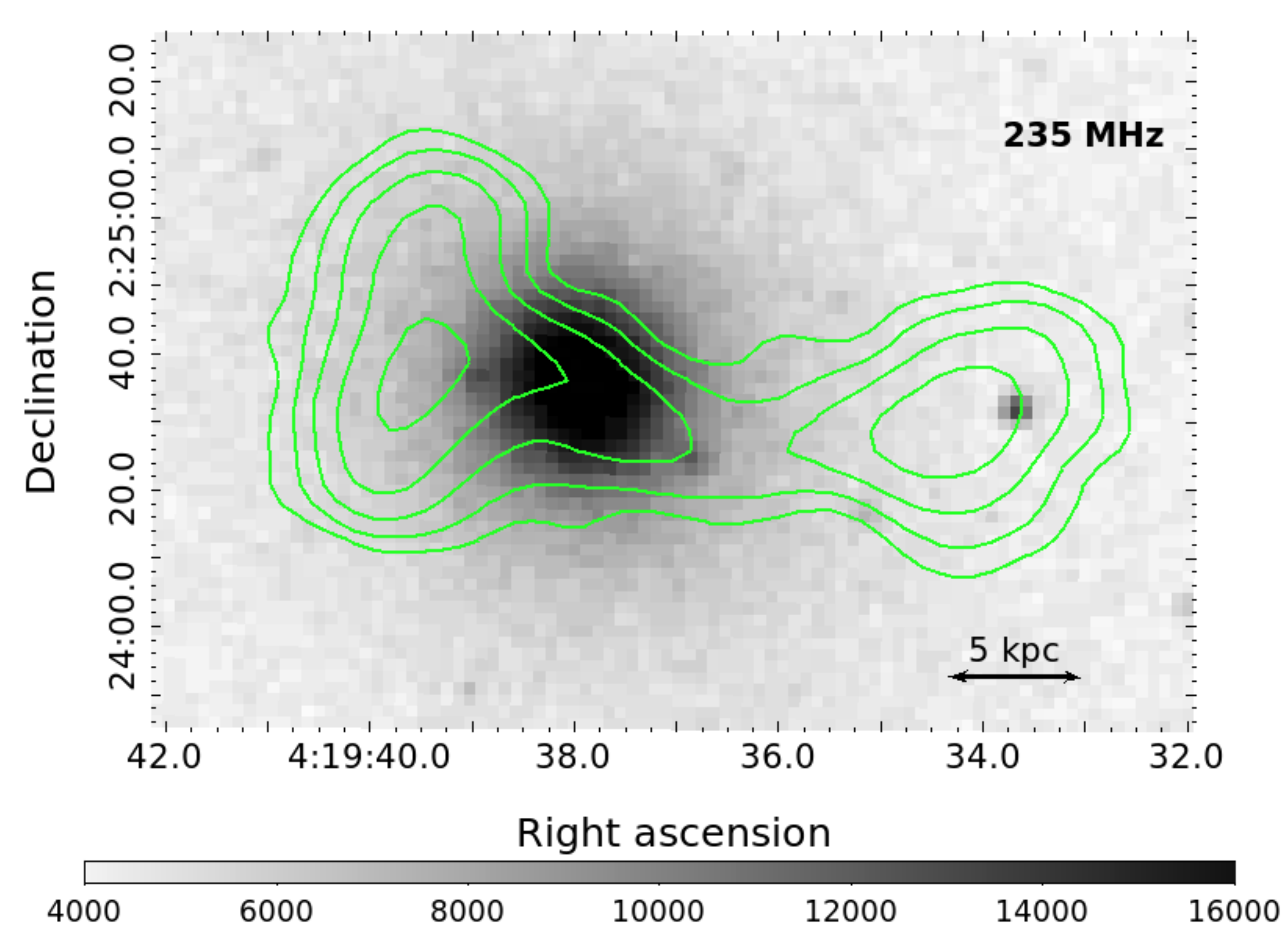}
\includegraphics[width=0.495\textwidth]{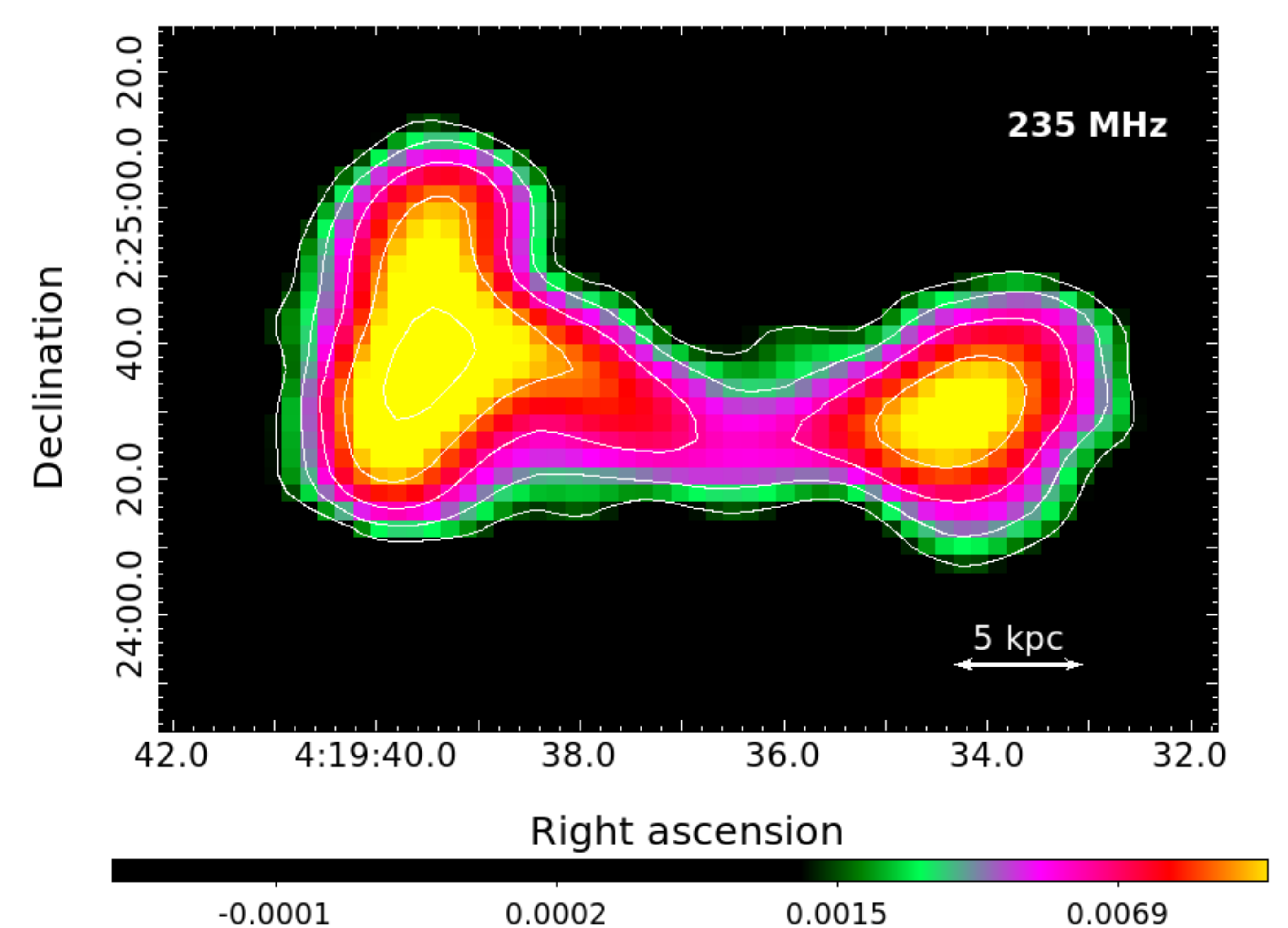}
\caption{\label{fig:radiocont235}\textit{Left:} GMRT 235~MHz radio contours overlaid on a DSS optical image of the galaxy. \textit{Right:} GMRT 235~MHz radio image and contours of the galaxy.  In both panels Right ascension is on the x-axis and Declination is on the y-axis. Contour levels start at 3$\sigma$ and rise in steps of a factor of 2, with the 1$\sigma$ rms noise being $\sim$0.4~mJy beam$^{-1}$ at this frequency.} 
\end{figure*}

Figures~\ref{fig:radiocont610} and \ref{fig:radiocont235} present the GMRT 610 and 235~MHz images of NGC~1550 at resolutions of $\sim$5.7$^{''}$$\times$ 4.0$^{''}$ and $\sim$14.3$^{''}$$\times$ 10.7$^{''}$ respectively. In both figures (left panels) the radio contours are overlaid on the red optical Digitized Sky Survey (DSS) images and at both frequencies we observe an asymmetric jet$-$lobe structure roughly aligned along the east$-$west axis. We note that the radio axis does not necessarily correspond to the optical galaxy's major or minor axis \citep[see, e.g.,][]{DaviesBirkinshaw86}.

We observe that the eastern lobe appears to be adjacent to the core at 235~MHz because of the lower resolution at this frequency, with the radio source having no clear indication of an eastern jet at 235~MHz. Assuming the source to be in the plane of the sky, with no significant projection effects, the eastern lobe is found to extend $\sim$12~kpc from the optical nucleus of the galaxy at this frequency whereas the western lobe and jet extend $\sim$2$\times$ farther on the opposite side, out to $\sim$21~kpc from the core. The western jet is clearly visible as a separate structure at both frequencies. At 610~MHz, the higher resolution image provides insight on several previously-unknown interesting features of the radio source: a) a small ($\sim$2~kpc in extent) eastern jet-like structure is visible and appears to be terminating at an elongated hotspot-like structure that is bending towards the south while being enclosed by the eastern lobe, b) the western jet presents a sharp bend or kink to the south $\sim$2~kpc west of the nucleus, before continuing relatively straight to the western lobe and c) the western jet appears to be fainter in the middle, at $\sim$4.5~kpc between the southern bend and the western lobe.  


Table~\ref{Sourcetabletotal} summarizes the properties (flux density, radio power and spectral index\footnote{spectral index $\alpha$ is defined as S$_\nu\propto\nu^\alpha$ where $\nu$ is frequency and S$_\nu$ is the flux at that frequency}) of the radio source as a whole and Table~\ref{Sourcetablecomp} presents the properties of its individual components (core, jets and lobes). The largest projected linear size of the source is 127\arcs/~32.6~kpc at 235~MHz. The eastern jet extends $\sim$4~kpc whereas the western counterpart appears to extend $\sim$10~kpc before it expands into the western lobe. The western lobe is fainter than that in the east and in projection is $\sim$36\% smaller at 235~MHz and $\sim$22\% smaller at 610~MHz.


The radio core is unresolved in our 610~MHz image. The angular size of the beam at this frequency is $\sim$ 1.46~kpc $\times$ 1.02~kpc, but the size of the core is most likely smaller than this and we may be overestimating its projected area shown in Table~\ref{Sourcetablecomp}. Examining Very Large Array (VLA) archival data (project code AE110) we find that the core is detected  as an unresolved point radio source at 4.9~GHz with a flux density of $\sim$0.94~mJy (rms 0.15 mJy/beam, $6''\times4''$ resolution), presenting no sign of an extended emission. VLA 8.4~GHz data (project code AB878) show no radio emission at the position of NGC~1550 (rms 0.18 mJy/beam, $\sim$0.8$''$ resolution). Examining also observations from the first epoch of the Very Large Array Sky Survey (VLASS), we find that a compact radio source of $\sim$0.5~mJy (rms$\sim$0.1~mJy) is barely detected at the core of NGC~1550 from the high resolution ($\sim$2.5$''$), $\sim$3~GHz (S-band) observations. It therefore seems likely that the core is unresolved in all these datasets, and we treat the estimates of its radio properties in Table~\ref{Sourcetablecomp} as upper limits.



\begin{figure}
    \centering
   \includegraphics[width=0.5\textwidth]{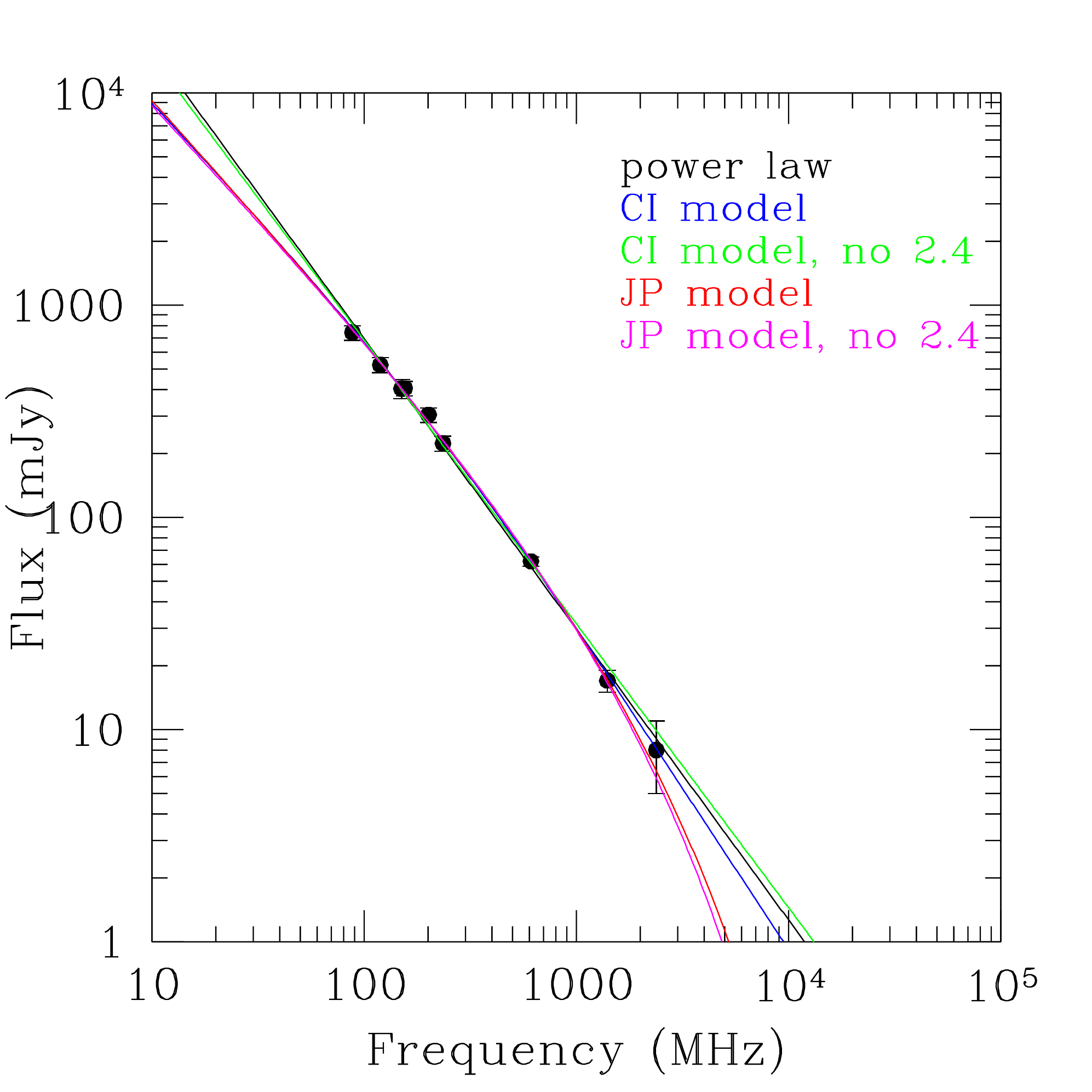}
    \caption{\label{fig:integratedfit2}Integrated spectral fitting of NGC~1550 using 3 different models: power-law, CI and JP. Table~\ref{tab:integratedfits} provides the  fit parameters for the different models used.}
\end{figure}

\begin{figure*}
    \centering
   \includegraphics[width=0.85\textwidth]{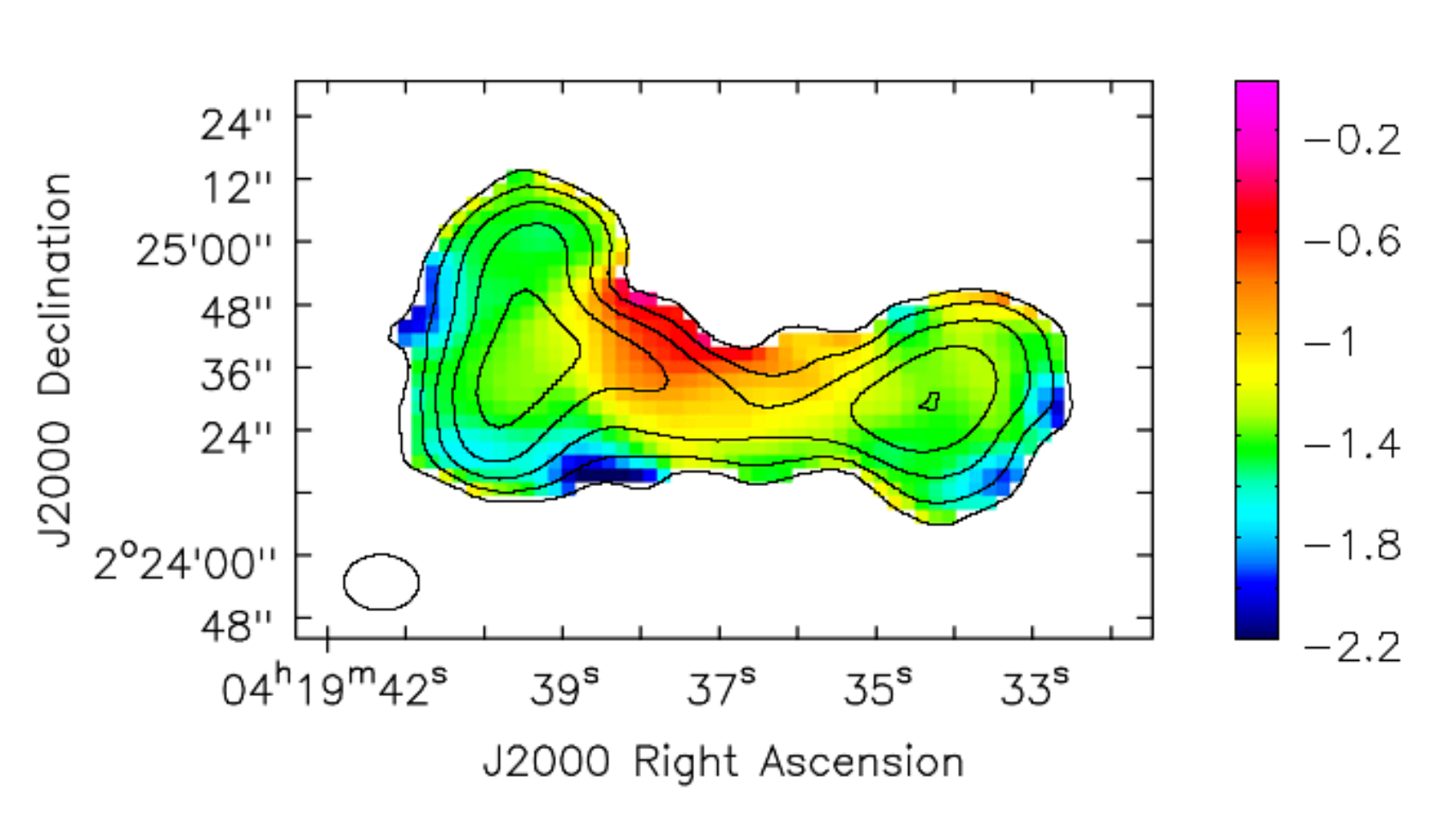}
    \caption{\label{fig:spix}Spectral index distribution map between 235 and 610~MHz based on images with matched resolution of $\sim$14$''$ HPBW. The contours represent the 235~MHz emission starting at the 3$\sigma$ level of significance. The beam size is marked by an ellipse at the bottom left. The typical uncertainty of spectral index is $\pm$0.04 with the values within one beam at the lobe edges being most likely unreliable.}
\end{figure*}


\subsection{Radio spectral analysis}
\label{sec:Radiospectra}

 To comprehend the nature of the various components of the radio emission in NGC~1550, we used the GMRT data analysed in this work to perform a spectral study of the radio source in the 235 $-$ 610~MHz frequency range. The GMRT data were also combined with data from the literature in order to estimate the integrated radio spectrum using several different frequencies in the range between 88~MHz $-$ 2.38~GHz. The flux densities are summarized in Table~\ref{Integratedspix}. The total spectrum of the source and the integrated model fits are discussed in \S~\ref{sec:Integratedspectra} with the spectral index
images of the source and the description of the spectral index profile distribution presented in \S~\ref{sec:spixprofile}. The calculation of the physical parameters and the radiative ages of the total source and its individual components are described in \S~\ref{sec:Radage} along with the description of the relevant spectral model fits.

\subsubsection{Integrated radio spectrum and spectral fits}
\label{sec:Integratedspectra}
The integrated radio spectrum of NGC~1550 was derived between 88~MHz and 2.38~GHz using flux density values from the literature (Table~\ref{Integratedspix}) along with our GMRT data at 235 and 610~MHz (Table~\ref{Sourcetabletotal}). The flux density measurements at 235 and 610~MHz were obtained using the task TVSTAT in AIPS, integrating within the 3$\sigma$ contour level at each frequency. The 150~MHz data were extracted from the TGSS Alternative Data release \citep[ADR;][]{Intemaetal17} catalog with the 150~MHz flux density value used for the integrated spectrum analysis not being the one reported (332$\pm$34~mJy) in the TGSS-ADR catalog, but integrated directly from the TGSS-ADR mosaic image. The flux density values at 88~MHz, 119~MHz, 155~MHz and 201~MHz were extracted from the wide-band observations of the GLEAM  \citep[GaLactic and Extra-Galactic All-Sky MWA;][]{HurleyWalkeretal17} survey, by also integrating the flux density directly on the images. In addition to these measurements, we include from NED\footnote{NASA Extragalactic Database (NED; https://ned.ipac.caltech.edu/)} the flux density at 1.4~GHz from \citet{Brownetal11} and the 2.38~GHz Arecibo single dish measurement from \citet{DresselCondon78}. 

Although the GLEAM survey and the 1.4~GHz measurement use a different flux density scale \citep{Baarsetal77} to that  used for TGSS-ADR survey and our analysis \citep{ScaifeHeald12} we note that the difference between the two scales is of the order of $\sim$3 per cent or less \citep{Perleyetal17}, which we consider negligible.



We performed three different radiative aging model fittings of the integrated spectrum over 9 frequencies between 88~MHz and 2.38~GHz using the Synage++ package \citep{Murgia01}. Figure~\ref{fig:integratedfit2} shows the total broad-band spectrum of NGC~1550 along with the spectral aging model fits. The details of the model fits are shown in Table~\ref{tab:integratedfits}. We find that the GMRT data points align well with the data from the literature with the total integrated radio spectral index being $\alpha_{88MHz}^{2380MHz}=-1.64\pm0.17$. The solid black line in Figure~\ref{fig:integratedfit2} indicates the model fit of a simple power-law for the source with $\alpha_{\mathrm{inj}}=-1.37_{-0.06}^{+0.01}$. We note that the simple power-law fit is in good agreement with the literature data and that similar steep spectral indices within uncertainties are also found in the narrower frequency ranges, (see Table~\ref{Sourcetabletotal}). The derived spectrum for the source is therefore steep and mainly dominated by its radio lobes. 




 A second model fit was performed using the continuous injection model (CI model; \citealt{Kardashev62}) where a continuous injection/flow of particles from the radio source's core is assumed. The third model fit was performed using this time the Jaffe \& Perola model (JP model; \citealt{JaffePerola73}) in which the particles originate from a single injection and the timescale for continuous isotropization of the electrons is assumed to be much shorter than the radiative timescale (the pitch angle of the particles is considered to be constant). The flux density measurement at the highest radio frequency available in the literature made by the Arecibo telescope at 2.38~GHz plays an important role in the determination of the source's break frequency, but its large uncertainty makes this task difficult. In order to map in detail the variation of the break frequency in respect to the models used, we performed two separate spectral fits for each of the CI and JP models, one that includes the 2.38~GHz data in the analysis and another one that excludes them (Figure~\ref{fig:integratedfit2}). We find that both the CI and JP models give a break for the spectrum of the radio source only if the 2.38~GHz data are included, but at very different frequencies. We note that the power-law fit gives no estimate of the break frequency for the radio spectrum. For the CI model with the 2.38~GHz data included in the analysis, we find a break frequency of $\nu_{\mathrm{break}}\sim$457~MHz with the one from JP giving a break at $\nu_{\mathrm{break}}\sim$4.6~GHz. Excluding the 2.38~GHz data we only get a break frequency for the JP model at $\nu_{\mathrm{break}}\sim$3.7~GHz. We find an order of magnitude discrepancy between CI and JP models on the $\nu_{\mathrm{break}}$. The inconsistency in the estimated break frequencies between the fitted CI and JP models can be attributed to the different natures of the two models and the large uncertainty at the highest frequency available from the single dish 2.38~GHz value. Hence, due to very little reliable spectral structure for a definite age to be extracted, we only mention as indicative the calculated age estimates in \ref{sec:Radage} and do not include the results from the integrated spectrum in the later discussion of source's age. More high frequency data for this source are required in order to discriminate between the two models.

\begin{figure}
    \centering
   \includegraphics[width=0.48\textwidth]{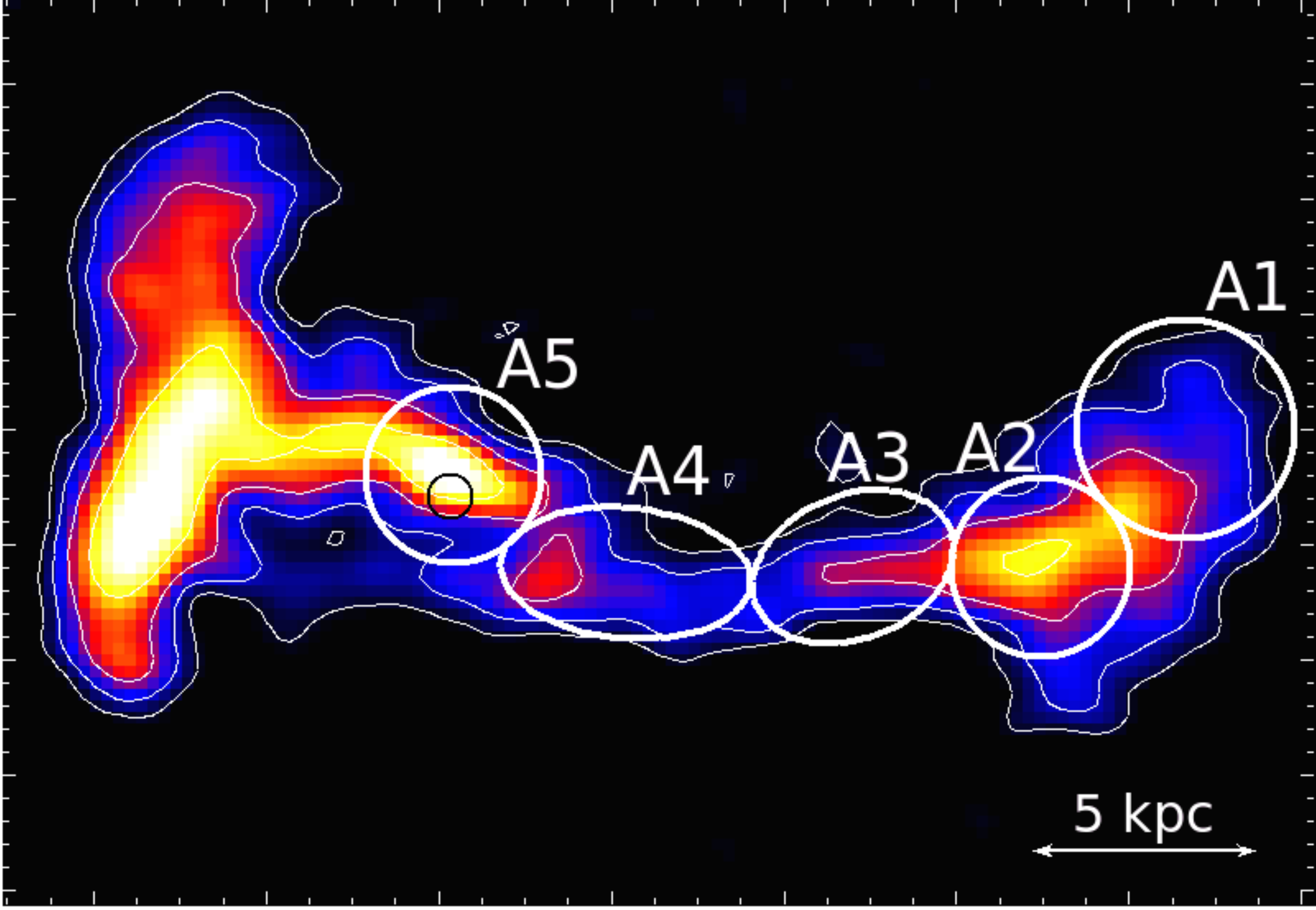}
    \caption{\label{fig:jetlobespixprofile}Regions used for point to point spectral profile analysis (A5-A1). The contours represent the emission at 610~MHz  (as for Figure~\ref{fig:radiocont610}).}
\end{figure}


\subsubsection{Radio spectral index profile analysis}
\label{sec:spixprofile}





We can use our GMRT data to constrain the nature of the bent jet morphology visible in the high resolution 610~MHz image, and the lobe structure enclosing the jets at 235~MHz. We initially made a new primary-beam-corrected 610~MHz image in order to match the \textsc{uv} range, cell size and restoring beam (14.28$''$ $\times$ 10.65$''$) of the 235~MHz image. As a consequence, the quality of the 610~MHz image used to derive the spectral index distribution is poorer than the original produced from the analysis presented in Figure~\ref{fig:radiocont610}. The spectral index map was then produced by combining the 235~MHz image with the matched 610~MHz image using the task `COMB' in AIPS. The flux density for each frequency in the spectral index image was clipped at the 3$\sigma$ level of significance. The spectral index image is shown in Figure~\ref{fig:spix}. 

Figure~\ref{fig:spix} shows that the core area presents a typical value for an AGN outburst $\approx-0.5$ with the area north of the core (orange/red color) having the flattest spectral index of the radio source. The spectral index in both lobes ranges between $\approx-1.2$ and $-1.8$ (see also Table~\ref{Sourcetablecomp}), with the spectral index of the radio emission in the lobes appearing to steepen as one moves away from the core. We note here that the steepest spectral index values in the lobes are observed in the areas where the 610~MHz radio emission is the faintest, just above the 3$\sigma$ level of significance at the original resolution ($\alpha\approx-1.9$; light/dark blue). The expected higher error of the spectral index in these areas means they should be treated with caution. The eastern jet presents a steep spectral index value of $\alpha_{235 MHz}^{610 MHz}=-1.22\pm0.04$, whereas the more extended western jet presents also a steep spectral index of $\alpha_{235 MHz}^{610 MHz}=-1.17\pm0.04$, which both indicate past activity. The larger and more visible west jet appears to have a spectral index that ranges between about $-1$ and  $-1.3$.

Both lobes show spectral indices $\alpha_{235 MHz}^{610 MHz}\approx-1.50\pm0.04$, indicating the presence of an older electron population than the one observed in the jets. The fact that the spectrum of the core is not flatter suggests that it most probably includes emission from an inner jet \citep[which might be expected to have a spectral index of about -0.7 to -0.8 as in, e.g., 3C66B,][]{HardcastleBW01} and/or the lobe, as steep-spectrum lobe emission is visible in its vicinity.

\begin{figure}
    \centering
   \includegraphics[width=0.5\textwidth]{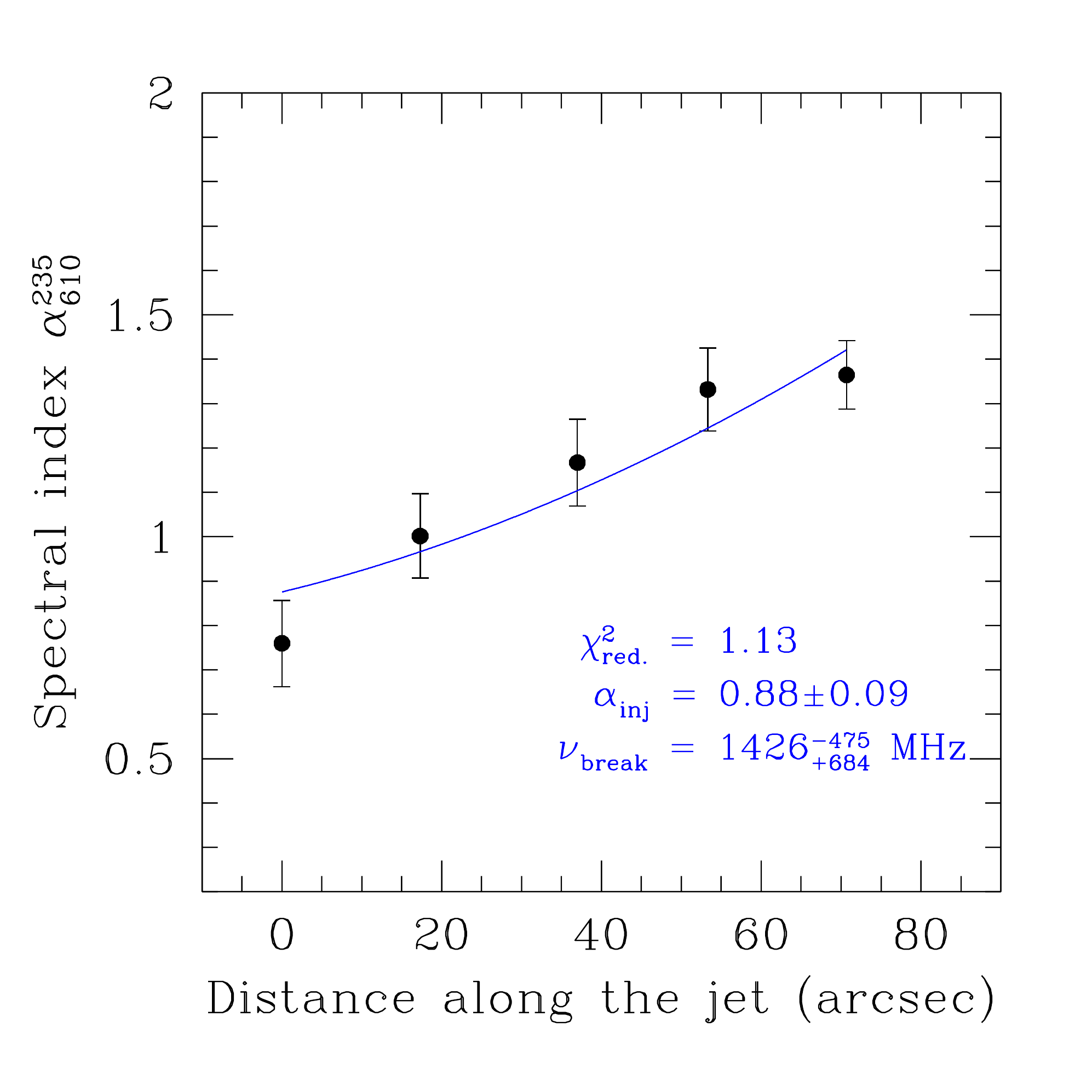}
    \caption{\label{fig:spixprofile}Spectral index profile along the western jet using the regions A5 to A1 and a JP model assuming that the break frequency scales as \textit{d}$^{-2}$ (\textit{d}$=$distance along the jet).}
\end{figure}

Whereas the integrated spectrum accounts for the source's emission in total, blending various areas of the source that may have different ages or origin, in the point-to-point spectral profile analysis we focus on extracting spectral information from specific regions trying not to mix up emission from distinct source components.
A point-to-point spectral index profile analysis was performed by fitting the observed spectral index trend along the source's axis, from the source's core towards the western lobe, following the methods used in \citet{Parmaetal07,Giacintuccietal08,Murgia03}. The Synage++ package has the ability to account for spectral indices between two frequencies, hence we modeled the trend in $\alpha_{235}^{610}$. Figure~\ref{fig:jetlobespixprofile} shows the circular and elliptical regions (A5 $-$ A1) used, based on the contours of the original highest resolution 610~MHz image. We mapped the trend starting from the region that most probably consists of the source's youngest plasma (A5, core), towards the region with the oldest plasma at the tip of the western lobe (A1) considering region A1 to be a rising plume of plasma without any jet input where buoyancy is the dominant effect. The regions were chosen to be larger than one beam in order that the spectral index measurements be independent. The radio images were matched in \textsc{uv}range, cellsize and resolution. Figure~\ref{fig:spixprofile} shows the derived $\alpha_{235}^{610}$ spectral index distribution in relation to the distance from the source's core.  We fitted a JP spectral aging model across regions A5 to A1 under the assumption that the expansion velocity of the source is constant and that the break frequency, $\nu_{\mathrm{break}}$, is $\propto d^{-2}$ where d is the distance from the central nucleus of the source \citep{Giacintuccietal08}. This argument implies that the radio emitting electrons are older at larger distances from the core, which is in accordance with the typical expansion assumption of lobe sources such as the western part of NGC~1550. The bent morphology of the eastern lobe indicates a strong interaction of some sort, hence an estimation of its spectral index trend would not be representative of the radiative history of the source and would not provide a trustworthy estimate for the source's age.

Figure~\ref{fig:spixprofile} shows the best such model fit to the observed spectral index data. We find that a JP model gives a $\nu_{\mathrm{break}}$ of 1426$^{+684}_{-475}$~MHz with $\alpha_{\mathrm{inj}}=-0.88\pm0.09$. The injected spectral index for the particles, $\alpha_{\mathrm{inj}}$, is less steep than the spectral index derived for both jets (see Table~\ref{Sourcetablecomp}) with the estimation of $\nu_{\mathrm{break}}$ coming with large uncertainties. The value for the $\nu_{\mathrm{break}}$ derived from the point-to-point spectral index trend analysis differs from the unreliable $\nu_{\mathrm{break}}$ values that we get from the fit of the integrated spectrum (see Table~\ref{tab:integratedfits}; more details in \S~\ref{sec:sourceage}).

\subsubsection{Radiative age and physical parameters}
\label{sec:Radage}

In order to estimate the physical parameters (e.g., minimum energy magnetic field, volume, particle and magnetic field energy densities) of the NGC~1550 radio source, we make a few specific assumptions. First, that the relativistic electrons and magnetic field energy densities are uniformly distributed over the total volume of the radio source, second that the magnetic field is constant throughout the lifetime of the radio source and third, that the total energy density from relativistic electrons and magnetic field is a minimum, which is similar to the electron and field energy densities being in equipartition (see also \citealt{Giacintuccietal08,Giacintuccietal12}). As in \citet{Kolokythasetal15}, we adopt a low energy cut-off Lorentz factor of $\gamma_{\mathrm{min}}=100$ in the energy distribution of the radiating electrons (which corresponds to $\approx50~\rm{MeV}$) and assume a tangled magnetic field along the line of sight. In addition, we note that our GMRT flux density measurement at 610~MHz was used (Table~\ref{Sourcetabletotal}) to estimate the source's radio luminosity, since the higher resolution at this frequency gives a more representative view of the morphology of the source.

\begin{figure}
    \centering
    \includegraphics[width=0.49\textwidth]{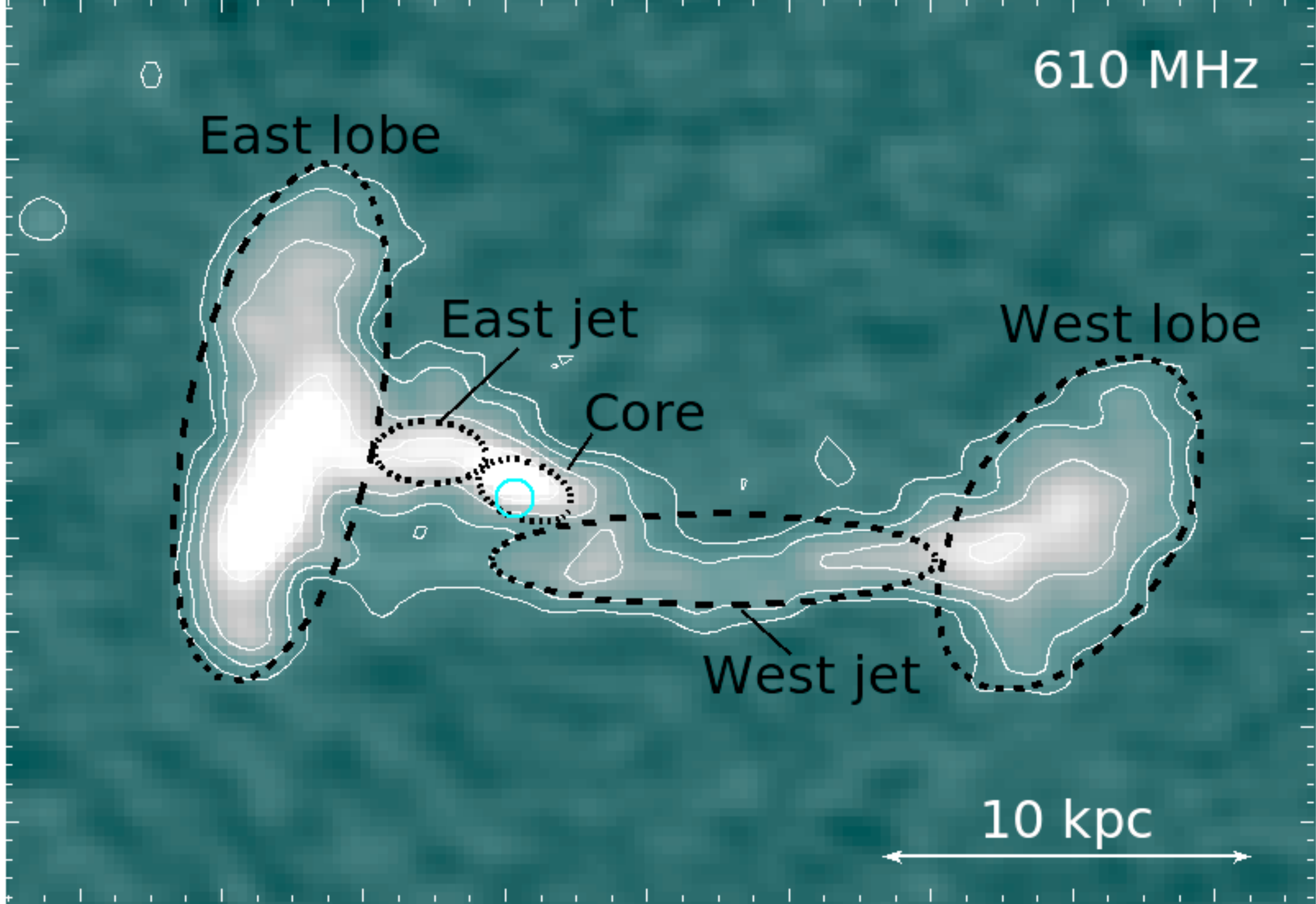}
    \caption{\label{fig:LobesC}Radio components of NGC~1550. The cyan circle indicates the position of the core with the grey scale representing the emission at 610~MHz  (as for Figure~\ref{fig:radiocont610}). The contours levels start at 3$\sigma$ level of significance. Table~\ref{Sourcetablecomp} gives the estimated properties for each of the radio components shown here.}
\end{figure}

We carefully select the regions that include every component at 610~MHz using the radio contours of the image at full resolution. For each region we obtained the integrated flux density at 610~MHz and estimated their projected areas. We note here the difficulty in defining the east jet component as it does not present a clear extension and is enclosed by the radio lobe (see Figure~\ref{fig:LobesC}).


For the source in total, and the regions of the components shown in Figure~\ref{fig:LobesC} that cover the corresponding areas of the NGC~1550 radio source, we assume ellipsoid symmetry about the major axis to estimate the volume in each jet region (prolate ellipsoid) and ellipsoid symmetry about the minor axis (oblate ellipsoid) to estimate the volume in the lobes. We estimate the minimum energy magnetic field using the equation \citep{WorrallBirkinshaw06}:

\begin{equation}
    B_{\rm min} = \left( \frac{(\alpha+1)C_1}{2C_2}\frac{(1+k)}{\phi V}L_\nu \nu^\alpha \frac{\gamma_{\rm max}^{1-2\alpha}-\gamma_{\rm min}^{1-2\alpha}}{(1-2\alpha)} \right)^{1/(3+\alpha)}, 
\end{equation}

\noindent where $\alpha$ is the spectral index (defined here as S$_\nu\propto\nu^{-\alpha}$), $C_1$ and $C_2$ are constants determined from synchrotron theory via eq (7) and (56) of the same paper, $k$ is the ratio of energy in non-radiating particles to that in electrons, $\phi$ is the filling factor of the relativistic plasma, $V$ is the volume of each region, $L_\nu$ is the radio luminosity at frequency $\nu$ and $\gamma_{\rm min}$ and $\gamma_{\rm max}$ are the limits on the electron energy distribution, expressed as Lorentz factors. We adopt $k$=0, $\phi$=1, $\gamma_{\rm min}$=100 and $\gamma_{\rm max}$=10$^5$.

We find minimum energy magnetic field of $B_{\mathrm{min}}\approx11\, \mathrm{\mu G}$ for the source in total, and  $B_{\mathrm{min}}$ values ranging from 11 to $16\, \mathrm{\mu G}$ for the different areas of the source. Table~\ref{tab:physicalparameters} shows the estimated minimum energy magnetic field $B_{\mathrm{min}}$ along with the derived parameters of the energy density in particles $u_{\mathrm{p}}$, the energy density in magnetic field $u_{\mathrm{B}}$ and volume V, for the source and its individual components. We find that the east jet presents the highest estimate for the $B_{\mathrm{min}}\approx16\, \mathrm{\mu G}$, whereas the west jet presents a lower magnetic field value of $B_{\mathrm{min}}\approx10\, \mathrm{\mu G}$. On the other hand, the east and the west lobe present estimates of $B_{\mathrm{min}}\approx13-15\, \mathrm{\mu G}$ which are slightly higher but similar to that of the total source. We note that the magnetic field values in the lobes and the source in total are typical of extended passive radio galaxies. 

Based on our ability to choose the size of the elliptical regions from the 610~MHz radio image, and the uncertainty of the flux scale for our data, we estimate the uncertainties on $B_{\rm min}$ to be 5-10 per cent with smaller uncertainties in the larger regions. We neglect the uncertainty contribution from spectral index, as it is small compared to the uncertainty on volume. 
However, the uncertainties associated with our assumptions, particularly the low-energy cutoff in the electron population, are considerably larger than the uncertainties on our measurements. We therefore emphasize that the values reported in Table~\ref{tab:physicalparameters} should be considered as approximate and indicative, rather than precise measurements.

Assuming the prevalence of radiative losses over expansion losses and ignoring re-acceleration mechanisms, using the $\nu_{\mathrm{break}}$ frequency \citep[e.g.,][]{MyersSpangler85,Giacintuccietal12} calculated from the point-to-point spectral index analysis in \S~\ref{sec:spixprofile}, the total radiative age, $t_{\rm rad}$, of the radio source in NGC~1550 and the radiative age, for every component individually can be estimated by means of the relation

\begin{equation}
    \centering
    t_{\rm rad}=1590\frac{B^{0.5}_{\rm min}}{B^{2}_{\rm min}+B^{2}_{\rm CMB}}[(1+\emph{z})\nu_{\rm break}]^{-0.5}\ {\rm Myr},
    \label{1}
\end{equation}

\noindent where $\nu_{\rm break}$ is expressed in GHz, and $B_{\mathrm{min}}$ and $B_{\rm CMB}$ in $\mathrm{\mu G}$ \citep{Parmaetal07}. $B_{\mathrm{min}}$ is the minimum energy magnetic field with $B_{\rm CMB}=3.2(1+\textit{z})^2$ being the equivalent magnetic field strength of the cosmic microwave background (CMB) radiation at redshift $z$. Hence, Equation~\ref{1} includes both synchrotron and inverse Compton losses.

From Equation~\ref{1} and the estimated mean minimum energy magnetic field for the overall source,  $B_{\mathrm{min}}\approx11\, \mathrm{\mu G}$, we find a total radiative age of $t_{\rm rad}\sim$33$^{-6}_{+8}$~Myr. We use the break frequency, $\nu_{\rm break}=1426$~MHz, derived from the point-to-point spectral profile analysis for the regions used in \S~\ref{sec:spixprofile} and find radiative ages of $t_{\rm rad}\approx$ 20 to 36~Myr in the various components of the radio source (see Table~\ref{tab:physicalparameters} for more details). We note that the JP model fits to the integrated spectrum (Table~\ref{tab:integratedfits}) provide a similar age estimate of $\approx$ 17 to 34~Myr, whereas the CI model gives a 2 $-$ 5 times larger estimate of $\approx$ 53 to 165~Myr. We adopt for the radio source in NGC~1550 the total radiative age of $t_{\rm rad}\sim$33~Myr based on the point-to-point spectral profile analysis.


\begin{table*} 
\caption{Estimation of the approximate physical parameters for the total source of NC~1550 and its components. For the total source age we used the break frequency of 1426 MHz that comes from the point-to-point spectral profile analysis JP model fit. \label{tab:physicalparameters}}
\begin{center}
\begin{tabular}{lccccc}
\hline 
&$B_{\rm min}$  &  $V_{\rm tot}$   & $t_{\rm rad}$  & $u_{\rm p}$ & $u_{\rm B}$\\ 
&($\mathrm{\mu G}$)  &  (kpc$^3$)   &   (Myr)    &   (10$^{-12}$ erg~cm$^{-3}$) & (10$^{-12}$ erg~cm$^{-3}$)  \\
\hline 
 Total source & 11.0  & 1040 & 33.4 & 4.1 & 4.8 \\
  East lobe  &  14.9  &  538 & 22.0 & 6.8 & 8.8 \\
  East jet   &   15.7 &  4.5 & 20.4 & 8.8 &  9.8 \\
  West jet   &  10.5  &  38 & 35.6 &  4.1 &  4.4 \\
  West lobe  &  13.4  &  297 & 25.5 &  5.6 &  7.1 \\
\hline
\end{tabular}
\end{center}
\end{table*}

\subsection{X-ray structures}
\label{sec:Xim}
\begin{figure*}
\includegraphics[width=0.49\textwidth,viewport=30 135 577 660]{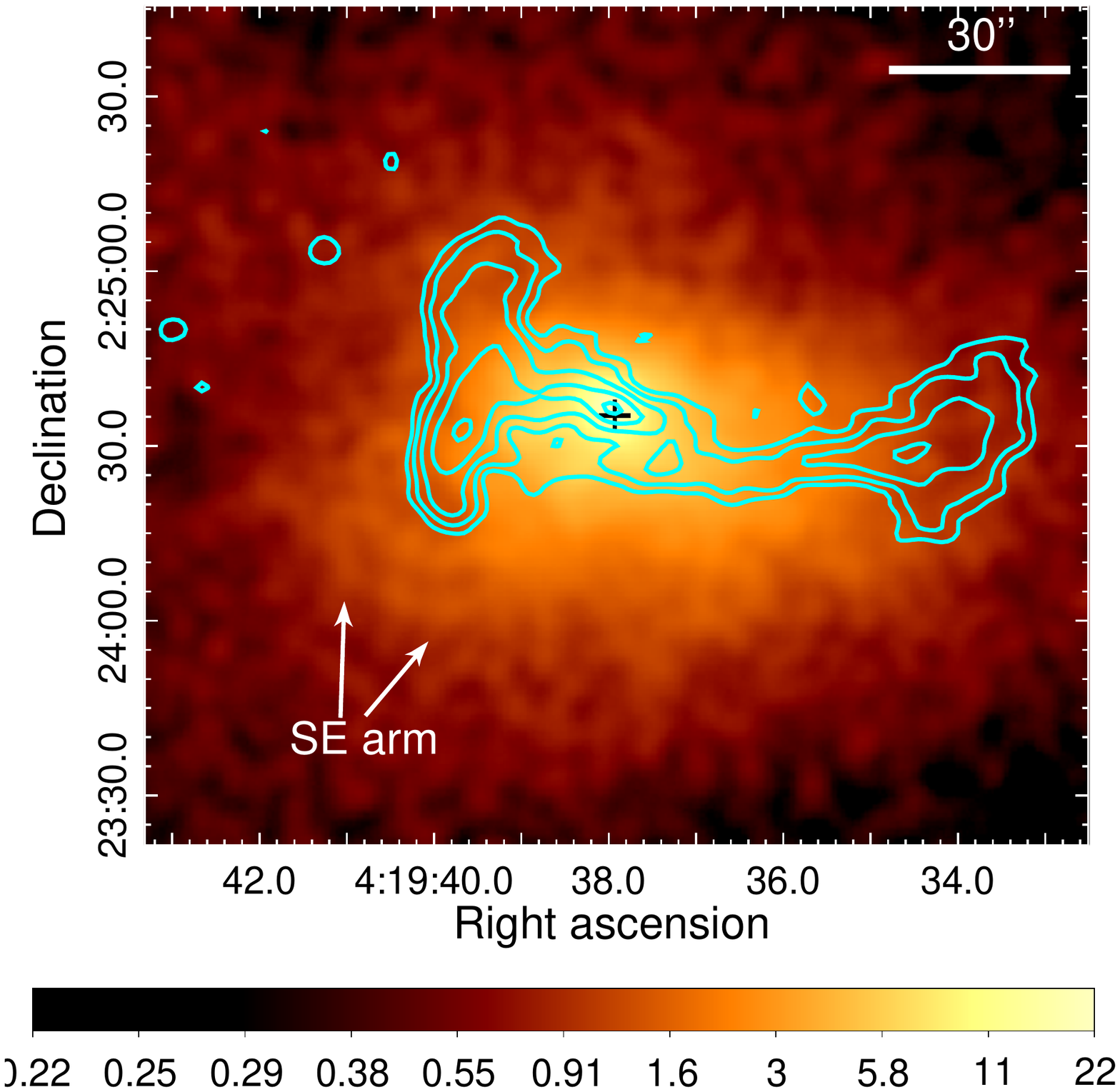}
\includegraphics[width=0.49\textwidth,viewport=30 135 577 660]{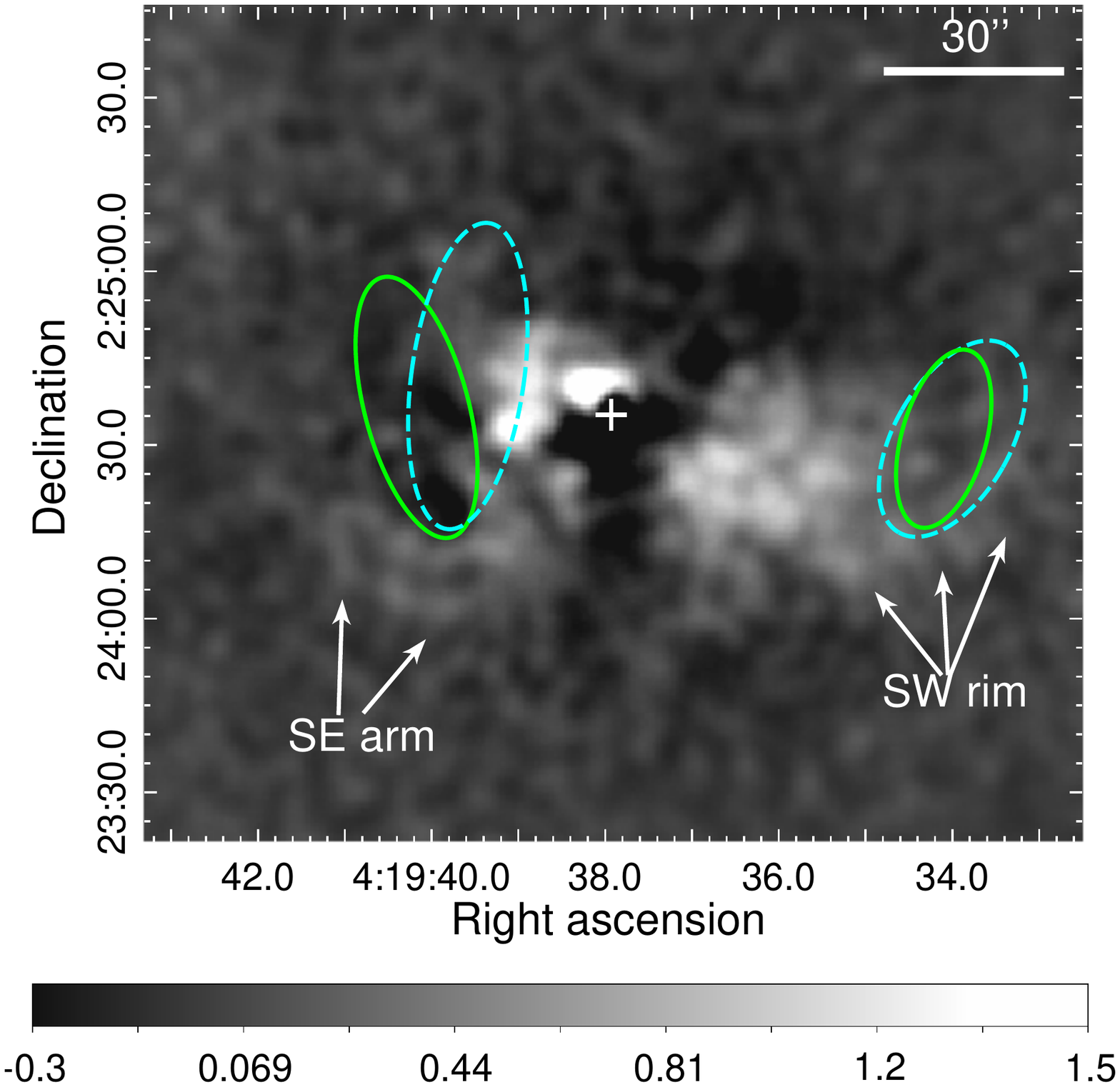}
\caption{\label{fig:Xim}\chandra\ 0.5-2~keV exposure corrected images of the core of NGC~1550, smoothed with a 4\arcs\ Gaussian and with the same angular scale and orientation. The \textit{left} panel shows the unmodified image with GMRT 610~MHz contours overlaid, starting at 3$\times$rms and increasing in steps of factor 2. The \textit{right} panel shows a residual map after subtraction of the best fitting surface brightness model. Ellipses mark the regions used for cavity enthalpy estimation, solid green for those selected from the X-ray data, dashed cyan for those based on the radio data. Structures discussed in the text are labelled, and the galaxy optical centroid is marked in both images by a cross. The scales of both images are in units of smoothed counts pixel$^{-1}$.}
\end{figure*}

Figure~\ref{fig:Xim} shows the GMRT 610~MHz contours overlaid on a smoothed, exposure corrected \chandra\ 0.5-2~keV image of the group core. The X-ray emission is clearly elongated along a roughly east-west axis. The X-ray emission peak is located at the optical centroid of the galaxy, but the brightest emission extends further to the west than to the east. The eastern margin of the bright emission roughly corresponds to the position of the east lobe, with the southern half of the lobe falling in a region of reduced surface brightness, bounded at its southern edge by a curved surface brightness enhancement (labelled SE arm). The extended emission to the west follows the line of the west jet, and there is a hint of a surface brightness depression at the position of the west lobe, partly surrounded by a slight enhancement.

To investigate these structures further, we carried out 2-dimensional surface brightness modelling, fitting a model consisting of an elliptical $\beta$-model plus a circular Gaussian to the overall surface brightness. The $\beta$-model was intended to subtract the overall group halo, while the Gaussian approximates any emission from the galaxy centre. All parameters (positions, normalizations, core radius, slope parameter $\beta$, FWHM of the Gaussian, and the ellipticity and position angle of the $\beta$-model) were allowed to fit freely. We note that this model was not fitted to derive physically meaningful parameters, but for completeness the fitted parameters are shown in Table~\ref{tab:SB}. As expected, the $\beta$-model approximates the extended X-ray halo of the group, while the Gaussian is compact. The centre of the $\beta$-model is $<$1\arcs\ from the optical centroid of NGC~1550, while the Gaussian component is offset by $\sim$2\arcs. The resulting model was subtracted from the image to create a residual map showing deviations from the smooth elliptical distribution. This is shown in Figure~\ref{fig:Xim}. 

\begin{table}
\caption{\label{tab:SB}Best-fitting parameter values and 1$\sigma$ uncertainties for the X-ray surface brightness model. Position angle is measured from due east.}
\begin{center}
\begin{tabular}{llc}
\hline
Component & Parameter & value \\
\hline
$\beta$-model & Centre RA & 04$^h$19$^m$37\fs 99$^{+0.88}_{-0.84}$ \\[+1mm]
& Centre Dec. & +02\degr 24\arcm 34\farcs 21$^{+0.66}_{-0.88}$  \\[+1mm]
& Core radius & 7.43$^{+7.94}_{-0.34}$\arcs \\[+1mm]
& $\beta$ & 0.39$^{+0.03}_{-0.01}$ \\
& Ellipticity & 0.19$\pm$0.05 \\ 
& Position angle & -9.0$\pm$4.6\degr \\
Gaussian & Centre RA & 04$^h$19$^m$37\fs 93$^{+0.50}_{-1.22}$\\[+1mm]
 & Centre Dec. & +02\degr 24\arcm 36\farcs 47$^{+1.72}_{-0.50}$\\
 & FWHM & 3.44$^{+3.82}_{-0.53}$\arcs \\
\hline
\end{tabular}
\end{center}
\end{table}

Because the emission in the inner region is elongated and asymmetrical, our simple elliptical model oversubtracts the emission immediately around the nucleus, to its south and particularly to the north; this is the cause of the darkest regions of the map. However, it also highlights the asymmetry of the bright emission in the group core. As well as the elongation of the emission to the west, the residuals show that the X-ray emission is more extended to the south of the optical centroid than to its north. We note that we attempted to add a second $\beta$-model component, to deal with any additional relaxed gas component (e.g., a cool core) but found that in the resulting best fit the component was unphysically elliptical. The fit was clearly driven by the extended emission to the west, but since this is neither symmetrical nor elliptical, the component was not able to accurately model it. The additional component made little difference to the residual image, and did not alter any of the structures significantly.

The elongated bright emission along the line of the west jet is visible as a bright residual, at the end of which we see a slight dip at the position of the west lobe, partly surrounded by the structure labelled "SW rim". The SE arm is clearly visible, and we see negative (dark) residuals in the east lobe. In both cases these structures may indicate cavities and partial rims of uplifted or compressed gas associated with each lobe. However, in the northern half of the east lobe, there is a plume-like positive residual curving from its inner west edge through to its northern tip. This could again indicate uplifted material, and may be seen in projection (i.e., it could be in front of or behind the lobe). Alternatively, it could indicate that the lobe structure is complex, with relativistic and thermal plasma structures mixing. Since plasma mixing would lead to depolarization of the radio emission, polarization observations could potentially determine the origin of the structure.

The elongated emission along the line of the jets is unlikely to be inverse-Compton emission arising from the radio source. It is anti-correlated with the lobes and broader than the 610~MHz jet.  Based on the radio properties of the west jet, we find the likely inverse-Compton 0.5-7~keV flux from cosmic microwave background photons to be $\sim$8.7$\times$10$^{-17}$~erg~cm$^{-2}$~s$^{-1}$ (flux density $\sim$5.5$\times$10$^{-12}$~Jy at 1~keV). This is equivalent to roughly one count over the duration of the ACIS-S observations. We expect the synchrotron self-Compton flux to be at least three orders of magnitude fainter, and inverse-Compton scattering of starlight or AGN photons would of course be concentrated in the galaxy centre, rather than producing the extended emission we observe.

\begin{figure}
    \centering
    \includegraphics[width=\columnwidth,viewport=36 135 577 660]{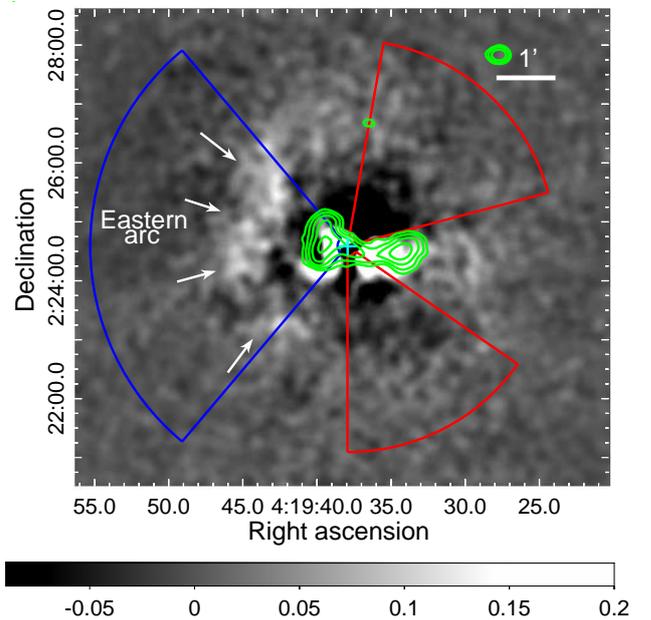}
    \caption{\label{fig:slosh}\chandra\ 0.5-2~keV residual image, smoothed with a 10\arcs\ Gaussian. The galaxy optical centroid is marked by a cross. The green contours represent the emission at 235~MHz (as for figure~\ref{fig:radiocont235}). Blue and red coloured wedges indicate angular (but not radial) ranges used to extract surface brightness and spectral profiles. The image scale is in smoothed counts pixel$^{-1}$, but note that since the smoothing scale differs from that of Figure~\ref{fig:Xim} the two are not comparable.}
\end{figure}

Figure~\ref{fig:slosh} shows the same residual map, smoothed on larger (10\arcs) scales. This blurs the structures associated with the radio lobes, but reveals an arc of positive (bright) residuals east of the core. The mid-point of this eastern arc is $\sim$1.75\arcm\ from the galaxy center, and the arc has a width $\sim$1\arcm.

\subsection{Cavity energetics}
\label{sec:Pcav}
We estimate the dimensions and position of the potential cavities in two ways, either from the X-ray surface brightness residual map, or from the 610~MHz radio image. In both cases we assume the cavities to be oblate ellipsoids. Cavity regions are marked in Figure~\ref{fig:Xim}.

\begin{figure}
\centering
\includegraphics[width=\columnwidth,viewport=35 450 555 740]{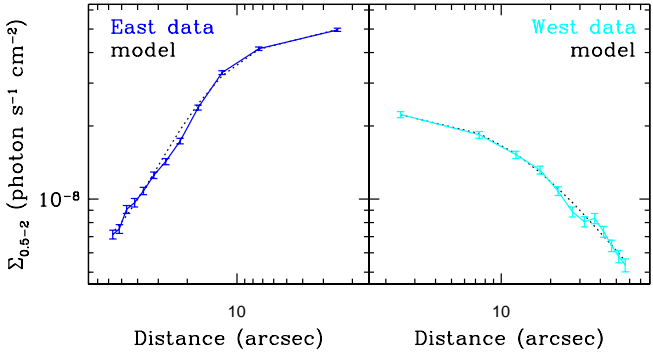}
\caption{\label{fig:cav_profs}Surface brightness profiles extracted across the two candidate cavities. Solid lanes and error bars show the data, dotted lines the best fitting $\beta$-models. Note that the horizontal axes only indicate distance along the profiles, not distance from the galaxy centre.}
\end{figure}

In a relaxed system, it is possible to estimate the significance of cavities from their deficit with respect to the best-fitting surface brightness model. However, the position of the cavities at the ends of the east-west excess means that the model of the group as a whole is not a good predictor of surface brightness at their positions. We therefore extracted profiles across each cavity using rectangular regions with E-W width 10 pixels ($\sim$5\arcs) and N-S height of either 50 (for the W candidate) or 90 pixels (for the E cavity), each profile having 12 regions crossing the cavity on an E-W axis. The region heights are chosen to approximate the cavity major axes. We fitted each of these profiles with a $\beta$-model, excluding the four central bins which contain the cavity and rims. The resulting fits are shown in Figure~\ref{fig:cav_profs}. In each case, two bins fall $>$1$\sigma$ below the model, while in the west profile, the bin immediately outside the cavity is $>$1$\sigma$ above the model. Taking the total deficit below the model in the two bins of the eastern profile, we find that it is 3.5$\sigma$ significant. For the western profile, the two-bin deficit is only 1.6$\sigma$ significant. In both cases the "rim" and "arm" features strongly suggest these are cavities; the SW rim corresponds to the 8$^{th}$ bin of the western profile, where we see a surface brightness excess. However, it seems that a deeper observation would be required to confirm the reality of the western candidate cavity.

\begin{figure}
\centering
\includegraphics[width=\columnwidth,viewport=20 250 570 770]{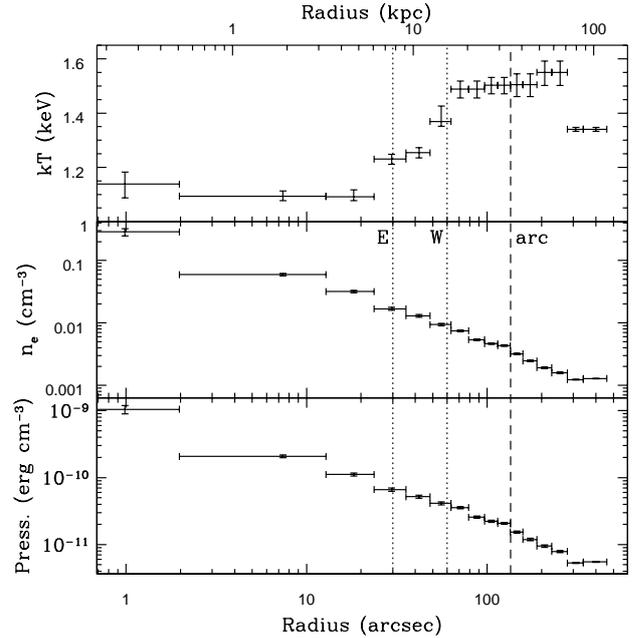}
\caption{Azimuthally averaged deprojected radial profiles of gas temperature, electron density and pressure in NGC~1550. The dotted lines mark the radii of the centres of the radio lobes, the dashed line the outer edge of the X-ray residual arc. }
\label{fig:Xprof}
\end{figure}

To estimate the kinetic power of the radio jets, we extracted spectra from a set of circular annuli centred on the galaxy optical centroid, with radii chosen to ensure a signal-to-noise ratio (S/N) of 125 in each annulus. We performed a deprojected spectral analysis using a \textsc{projct*phabs*apec} model in \textsc{Xspec}. The resulting entropy and pressure profiles are similar to those of \citet{Lakhchauraetal18}. Azimuthally averaged temperature (kT), electron density (n$_{\rm e}$) and pressure profiles of the IGM are shown in Figure~\ref{fig:Xprof}, where pressure is defined as 2n$_{\rm e}$kT.  While the western candidate cavity is not significantly detected, and we cannot be certain of the precise size of either cavity, the negative residuals and the SW rim and SE arm structures support the idea that the lobes have driven the IGM gas out of the majority of their volume. We estimate the enthalpy of each cavity as 4$PV$, where $V$ is the volume and $P$ the IGM pressure at the radius of the midpoint of the cavity. Cavity dimensions and enthalpies, estimated for both X-ray and radio-defined cavity sizes, are shown in Table~\ref{tab:Pcav}. 

\begin{table*}
\caption{\label{tab:Pcav}Dimensions, enthalpies, dynamical age estimates and jet power estimates for cavities at the location of the radio lobes. Cavity semi-major and -minor axes are given as r$_{\rm maj}$ and r$_{\rm min}$, and their distance from the galaxy nucleus by R.}
\begin{center}
\begin{tabular}{lccccccccccc}
\hline
Cavity & r$_{\rm maj}$ & r$_{\rm min}$ & R & Pressure & 4$PV$ & \multicolumn{3}{c}{cavity age} & \multicolumn{3}{c}{Jet power} \\
 & & & & & & t$_{\rm sonic}$ & t$_{\rm buoy}$ & t$_{\rm refill}$ & P$_{\rm sonic}$ & P$_{\rm buoy}$ & P$_{\rm refill}$ \\
       & (arcsec)   & (arcsec)   & (arcsec) & (10$^{-11}$~erg~cm$^{-3}$) & (10$^{57}$~erg) & \multicolumn{3}{c}{(Myr)} & \multicolumn{3}{c}{(10$^{42}$~erg~s$^{-1}$)}\\
\hline
\multicolumn{3}{l}{\textit{Cavity size determined from the X-ray}}\\
East & 23.2 & 9.0 & 35 & 6.62$^{+0.68}_{-0.34}$ & 2.68$^{+4.28}_{-0.79}$ & 12.8 & 27.9 & 37.0 & 6.64$^{+10.61}_{-1.96}$ & 3.05$^{+4.87}_{-0.90}$ & 2.29$^{+3.66}_{-0.68}$ \\
West & 15.9 & 7.3 & 57 & 4.00$^{+0.28}_{-0.14}$ & 0.80$^{+0.98}_{-0.27}$ & 19.8 & 52.6 & 33.0 & 1.28$^{+1.56}_{-0.43}$ & 0.48$^{+0.59}_{-0.16}$ & 0.77$^{+0.94}_{-0.26}$\\
\multicolumn{3}{l}{\textit{Cavity size determined from the radio}}\\
East & 26.7 & 9.7  & 30 & 6.62$^{+0.68}_{-0.34}$ & 3.82$^{+6.76}_{-2.80}$ & 11.0 & 21.3 & 36.4 & 11.05$^{+19.56}_{-2.93}$ & 5.69$^{+10.06}_{-1.51}$ & 3.64$^{+5.89}_{-0.89}$ \\[+1mm]
West & 20.8 & 10.0 & 60 & 4.00$^{+0.28}_{-0.14}$
& 1.20$^{+1.07}_{-0.30}$ & 27.1 & 48.6 & 37.5 & 1.82$^{+1.62}_{-0.46}$ & 0.78$^{+0.70}_{-0.20}$ & 1.01$^{+0.90}_{-0.26}$ \\
\hline
\end{tabular}
\end{center}
\end{table*}

We estimate the uncertainty on the cavity enthalpy, and three estimates of the cavity ages (buoyant rise time, sonic timescale, and refill time) following the approach outlined in \citet{OSullivanetal11b}. The jet power is defined as the enthalpy divided by the cavity age estimates, and the resulting values are also shown in Table~\ref{tab:Pcav}. The unabsorbed deprojected bolometric X-ray luminosity of the IGM within 79\arcs\ radius of the nucleus (equivalent to the distance to the tip of the west lobe) is [1.81$\pm$0.02]$\times$10$^{42}$~erg~s$^{-1}$. This is somewhat greater than the jet power estimates for the west lobe, but less than those for the east lobe. As is the case for many other group and cluster-central radio galaxies, this suggests that the AGN of NGC~1550 is capable of balancing radiative losses from the IGM and preventing excessive cooling, providing that the AGN jets are efficiently coupled to the IGM gas.

The strong asymmetry of the radio source, with the west lobe twice as far from the galaxy as the east lobe, also suggests that the development of the lobes has not been governed by simple buoyancy forces.

\section{Discussion}

The correlated radio and X-ray structures in the core of NGC~1550 demonstrate the impact of the AGN jets on the hot IGM. Both lobes appear to be associated with cavities, though we cannot confirm the presence of a western cavity. We also see rims or arms of gas at the edges of the lobes, suggesting that they have either uplifted material from deeper in the core, or compressed the surrounding gas as they expanded. Given the steep spectral index of the lobes, uplift seems the more likely explanation, since if the lobes are old they are likely to be in (or close to) pressure equilibrium with their surroundings, and any compressed gas features will have long since dispersed.

However, the asymmetry of the radio source, with the west jet roughly twice as long as its eastern counterpart, and the bend in the west jet, suggest that uplift is not the only physical process responsible for the observed morphology. A persistent pressure differential between the two sides of the core could provide an explanation. If the east jet were expanding into a higher pressure region it might be confined at smaller radii while the west jet would be able to expand further. However, there is no obvious cause for such a long-term pressure differential; indeed the asymmetric X-ray morphology suggests that, in the absence of other factors, the east jet had a shorter distance to expand before exiting the densest part of the core. The radio morphology is therefore the opposite of what would be expected in this scenario.


\subsection{Sloshing}
The arc-shaped structure in the X-ray residual map may hold the key to explaining the morphology of NGC~1550. The arc resembles structures observed in simulations of sloshing, the periodic oscillation of a group or cluster core set in motion by the tidal forces of a minor merger or flyby encounter. The oscillations occur in the plane of the orbit of the infalling perturber. When this plane is close to the plane of the sky, the sloshing produces a characteristic spiral pattern in surface brightness residuals, temperature and abundances, as cool, enriched gas is left behind by the motion of the core and warmer gas is drawn inwards \citep[e.g.,][]{Roedigeretal12,Gastaldelloetal13,Ghizzardietal13,Suetal17}. However, when the plane is orthogonal to the plane of the sky, our line of sight passes through the spiral, producing alternating positive and negative arc-shaped residuals to either side of the core as in, e.g., the galaxy group NGC~5044 \citep{OSullivanetal14a} or galaxy clusters such as Abell~1795 \citep{Markevitchetal01,Ehlertetal14}, Abell~2219 \citep{Canningetal17} or Abell~1664 \citep{Calzadillaetal19}.

\begin{figure}
    \centering
    \includegraphics[width=\columnwidth,viewport=36 152 577 640]{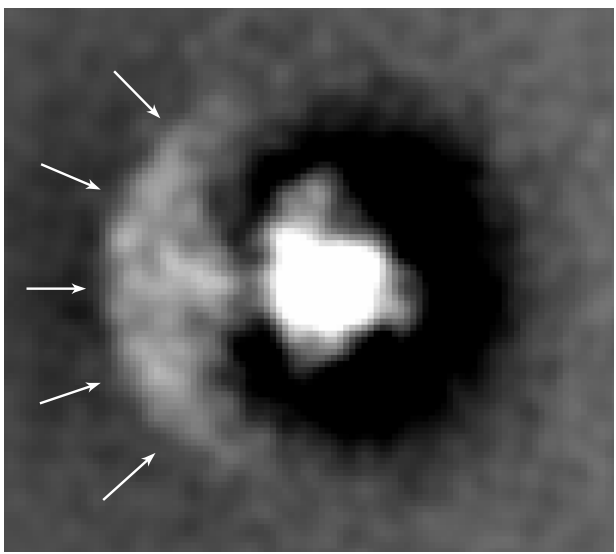}
    \caption{Residual image of a sloshing galaxy cluster, derived from a simulation in the Galaxy Cluster Merger Catalog. The sloshing is occurring in a plane orthogonal to the plane of the sky, and is driven by the infall of a 5:1 mass ratio gasless perturber. Note the bright arc east of the core, indicated by arrows, marking the outer edge of the sloshing cold front.}
    \label{fig:sim}
\end{figure}

The Galaxy Cluster Merger Catalog\footnote{http://gcmc.hub.yt} \citep{ZuHoneetal18} provides a set of simulated minor mergers designed to explore the effects of sloshing on galaxy clusters. The simulations assume a massive primary cluster (M$_{200}$=10$^{15}$~M$_\odot$, where M$_{200}$ is defined as the total mass of the system within a spherical volume whose interior density is 200 times the critical density of the universe) but we are only interested in the shapes of structures produced by the sloshing motions, which would be the same in a lower mass system. We examined a set of simulations for a 5:1 mass ratio merger with an impact parameter of 500~kpc, a gas-free perturber and no viscosity. The catalog provides simulated X-ray images at various stages of the merger. We applied the surface brightness model subtraction described in Section~\ref{sec:Xim} to these images, creating residual maps, one of which is shown in Figure~\ref{fig:sim}. Outside the bright core, the strongest feature in the image is a bright arc east of the core, similar to the arc we observe in NGC~1550. This arc consists of cold, enriched gas drawn out of the cluster core by the sloshing motion, with its outer boundary marking a cold front in the gas. The similarity to the observed structure in NGC~1550 strongly suggests that the group is also sloshing.

It should however be noted that in NGC~1550 the sloshing is unlikely to be perfectly orthogonal to the plane of the sky, and that the AGN jets will also have impacted the gas. We should not expect the comparison with the simulated cluster to be exact; the observed system is of course more complex than the simulation.

\begin{figure}
\centering
\includegraphics[width=\columnwidth,viewport=35 70 555 730]{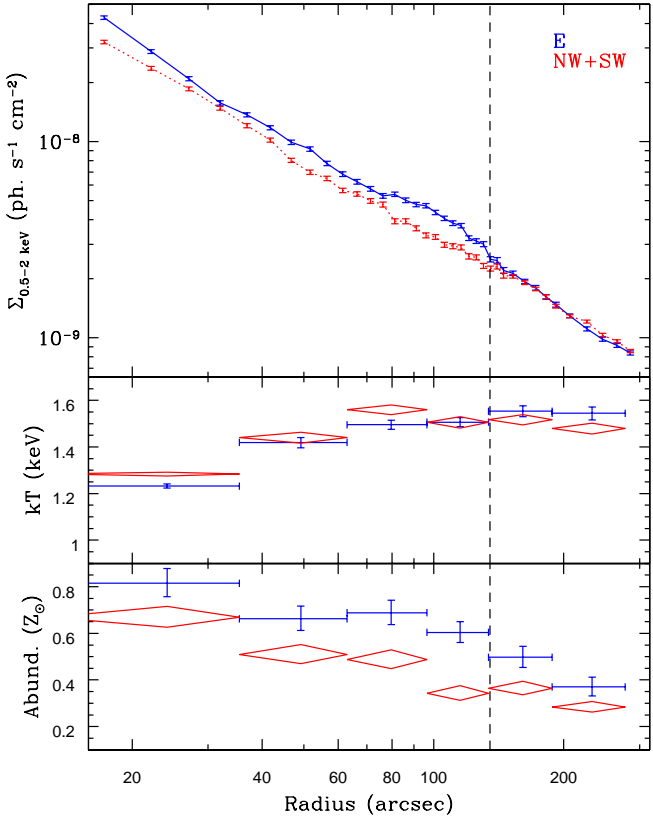}
\caption{\label{fig:SB}Radial profiles of 0.5-2~keV surface brightness, projected temperature, and projected abundance, extracted from sectors extending east (blue solid line, crosses) and a southwest+northwest (red dotted line, diamonds), as shown in Figure~\ref{fig:slosh}. Error bars and diamonds indicate 1$\sigma$ uncertainties. The dashed vertical line marks 135\arcs\ from the galaxy optical centroid.}
\end{figure}

Figure~\ref{fig:SB} shows 0.5-2~keV radial surface brightness profiles for NGC~1550, one extracted in a sector of opening angle 100\degr\ extending due east, the other a combination of two sectors extending roughly northwest (15-80\degr, measured from due west) and southwest (270-325\degr). Figure~\ref{fig:slosh} shows the angular extent of the regions. The eastern sector was chosen to cross the eastern arc identified in the residuals to the surface brightness fit, the southwest and northwest sectors to avoid any of the structures in the group core. There is a clear excess in the eastern profile compared to the combined southwest+northwest profile, centred at around 1.75\arcm, corresponding to the eastern arc, confirming the significance of this structure. The surface brightness decline at the outer edge of the excess ($\sim$135\arcs) is quite steep, consistent with a change in gas properties across a sloshing front.

We would expect material transported outward from the group core by sloshing motions to be cooler and/or enriched compared to its surroundings. To test this, we extracted spectra in the same sectors used for the surface brightness profile and measured the projected temperature and abundance, fitting a \textsc{phabs}*\textsc{apec} model in \textsc{Xspec}. The resulting profiles are shown in Figure~\ref{fig:SB}. We find that the temperature profiles are similar and relatively flat in the range $\sim$60-280\arcs, as is also the case for the deprojected azimuthally averaged temperature profile of the group as a whole shown in Figure~\ref{fig:Xprof}. The temperature in the E profile peaks immediately outside the E arc, while that in the NW+SW profile peaks at $\sim$80\arcs\ and appears to decline slightly outside that. The abundance profiles show larger differences, with the E profiles showing higher abundances than the NW+SW profile at all radii. The difference is greatest in the fifth bin of the profile, which corresponds to the region immediately inside the E arc. However, the abundances are still higher in the E than in the NW+SW profile, even at radii beyond the E arc. Deprojected profiles in these sectors show the same trends, but with poorer spatial resolution, as we have to tie abundances between bins to achieve a stable profile.

The change in temperature and abundance from the bin inside the E arc to the bin outside it means that, in the absence of any change in gas density, we would expect the 0.5-2~keV flux to drop by $\sim$10 per cent across the boundary. The surface brightness profile actually declines by $\sim$16 per cent suggesting that there is also a decline in density. However, the uncertainty on the surface brightness jump is $\pm$4 per cent so while the change in overall surface brightness is $\sim$4$\sigma$ significant, deeper data would be needed to formally confirm a density jump. The temperature increase across the boundary is $\sim$3 per cent meaning that the pressure difference across the boundary is, within uncertainties, consistent with zero, as expected for a sloshing front.

Overall, we consider these results to be consistent with sloshing. The lack of a strong temperature jump is understandable given the flat temperature profile at the radius of the E arc, while the location of the largest abundance excess inside the arc is what we would expect from sloshing. However, the presence of a smaller abundance excess outside the arc suggests that enriched gas may have been transported to larger radii. This could indicate that sloshing has caused motions beyond the E arc, and that there are other sloshing fronts to be found at larger radii. Alternatively, it may be a sign that earlier periods of AGN activity have left the system with an asymmetric abundance distribution.

Sloshing motions may also explain the asymmetry of the radio source. If NGC~1550 is in motion relative to the gas around it, the jets and lobes may be bent back, appearing to have unequal lengths in projection. For example, if NGC~1550 were moving toward us and to the east, a significant fraction of the length of the east jet might be along the line of sight, while most of the west jet's extent would be in the plane of the sky. However, given the flattened shape of the east lobe, it seems likely we are observing it from the side, and that the original axis of the east jet was therefore close to the plane of the sky. This suggests that much of the motion is to the east, and that the jets were initially of roughly equal length, but the galaxy has since moved closer to the east lobe. Relative motions between the jets, lobes, nucleus and surrounding IGM could also potentially explain the bend in the west jet.

We can make an order-of-magnitude estimate of the sloshing timescale following the approach described by \citet{Churazovetal03}, calculating the Brunt-V\"{a}is\"{a}l\"{a} frequency at the radius of the sloshing front $r$ \citep{BalbusSoker90}:

\begin{equation}
    \omega_{\rm BV}=\Omega_{\rm K}\sqrt{\frac{1}{\gamma}\frac{d{\rm ln}K}{d{\rm ln}r}},
\end{equation}

\noindent where $K$=kT/n$_e^{2/3}$ is the entropy of the gas (as usually defined in astrophysics, the true entropy being proportional to ln$K$), $\Omega_{\rm K}$=$\sqrt{GM/r^3}$ is the Keplerian frequency, and $\gamma$=5/3. We take the radius of the front to be $r$=135\arcs\ (34.7~kpc) and determine the gas entropy and total mass ($M$) profile from the deprojection analysis. The sloshing period is given by $P$=2$\pi$/$\omega_{\rm BV}$ but we might expect to see a front after only half an oscillation, hence the sloshing timescale may be $P$ or $P$/2. On this basis, we estimate the age of the sloshing front to be 40-80~Myr.

\subsection{Uplift}
\label{sec:uplift}
The asymmetry of the X-ray emission in the group could in principle arise either from uplift of gas by the radio lobes, or from gas being drawn out of the core by the sloshing. Evidence of uplift by radio galaxies has been observed in many clusters, both in the hot ICM  \citep[e.g.,][]{Sandersetal04,Simionescuetal09,Kirkpatricketal09,Kirkpatricketal11,OSullivanetal11a} and in denser material associated with the cooling flows that fuel the AGN \citep[e.g.,][]{Salomeetal11,Russelletal16,McNamaraetal16,Tremblayetal18,Russelletal19}. The lobes can potentially uplift a mass of gas equal to that which they displace, but simulations suggest that the process would only be $\sim$50\% efficient \citep{Popeetal10}. Assuming the lobes to be in the plane of the sky, and adopting the volumes and deprojected profiles used in the calculation of cavity enthalpy, we estimate that the east lobe could uplift $\sim$2.8$\times$10$^7$\Msol\ and the west lobe $\sim$0.8$\times$10$^7$\Msol\ of gas. 

The region of enhanced X-ray emission along the west jet is $\sim$6.5~kpc in width and extends $\sim$12.5~kpc from the nucleus to the base of the west lobe. Excluding a 3.25~kpc radius core around the nucleus, we can approximate the remaining volume along the jet as a cylinder, which we estimate contains $\sim$2.1$\times$10$^8$\Msol\ of gas (assuming the gas to be evenly distributed). The west lobe could only uplift $\sim$13\% of this mass.

The properties of the gas along the west jet differ from the surrounding regions. Along the jet, the projected temperature is 1.22$\pm$0.01~keV and the abundance is 0.68$_{-0.03}^{+0.05}$\Zsol, compared to 1.40$\pm$0.03~keV and 0.51$\pm$0.05\Zsol\ in the gas north and south of the jet. We therefore expect a difference in emissivity. Based on an APEC thermal plasma model with Galactic absorption, we find that the gas along the jet should be $\sim$1.4 times brighter than the gas to its south. The actual difference in surface brightness is a factor $\sim$1.26, though this value may be affected by structure along the line of sight. The gas along the jet is therefore brighter primarily because it is cooler and more enriched, consistent with this material originating closer to the core. 

We simulated the effect of uplift by creating a spectrum where 87\% of the emission comes from 1.5~keV, 0.5\Zsol\ gas and 13\% comes from cooler gas with properties like those found in the core of NGC~1550. Our deprojection analysis showed that in the central 2\arcs\ of NGC~1550 the temperature is  1.14$^{+0.04}_{-0.05}$~keV and abundance is 0.85$^{+0.40}_{-0.25}$\Zsol. The simulated spectra were then fitted with a single temperature absorbed APEC model. We find that even with only this small amount of uplifted gas, it is possible to reproduce the apparent enrichment along the jet, though slightly cooler gas than we observe in the core would be required to reproduce the temperature. Given that the core may have been cooler before the uplift occurred, and that uplifted gas may cool adiabatically as it is drawn out into lower-pressure regions, it is therefore plausible that uplift could be responsible for the structure we see along the west jet. However, it cannot explain the overall asymmetry of the X-ray emission. The west jet is less able to lift gas than its eastern counterpart, but appears to have travelled further and is associated with a greater extension of the X-ray surface brightness. This is the opposite of what we would expect if uplift were the only process at work. It therefore seems certain that gas motions caused by sloshing have also played a role in shaping these structures.

\subsection{Radio source age and development}
\label{sec:sourceage}

The radiative age estimates derived from the radio observations ($\sim$22-25~Myr for the lobes, $\sim$33~Myr for the source as a whole) are shorter than, but comparable to the shortest timescale estimate for the sloshing motions ($\sim$40~Myr). The radiative age of the source as a whole is in reasonable agreement with the range of dynamical estimates for the cavities, at least for the refill ($\sim$33-38~Myr) and buoyancy ($\sim$21-53~Myr) timescales. Comparison between the radiative and dynamical timescale estimates for the individual lobes is more complex. The radiative age of the west lobe ($\sim$25~Myr) is consistent with the sonic timescale ($\sim$20-27~Myr, depending on whether X-ray depression or radio lobe size is used), but is shorter than the buoyant or refill timescales. For the east lobe, the radiative age ($\sim$22~Myr) is closest to the buoyant timescales ($\sim$21-28~Myr), longer than the sonic timescale and shorter than the refill timescale. The lack of sharp features associated with the lobes indicates that their motions are subsonic, therefore the sonic timescales are likely underestimates of the true age. If sloshing has affected the development of the source, we would also expect the buoyant timescales to be affected, as the distance of the lobes from the AGN would no longer be determined primarily by the buoyant rise time. We can therefore only suggest that the range of the dynamical timescales is comparable to, but may be somewhat longer than, the radiative age of the lobes.

The morphology of the radio source, with its asymmetry and the kink in the west jet, strongly suggests that it has been affected by the motions of the surrounding gas. This is consistent with the source being comparable in age to the sloshing motions. It seems likely that sloshing would only have a limited effect on a radio source during the period of jet activity, as young radio jets and lobes will have internal pressures significantly greater than the additional external pressure from the sloshing motions. It is clear that sloshing motions can, and do, impact the location and shape of older radio lobes within galaxy groups. Examples include NGC~5044, in which sloshing appears to have bent an old radio jet and separated it from its associated lobe \citep{OSullivanetal14a}, and NGC~507 where the shape of the eastern lobes traces the inner edge of an X-ray front whose origin is likely sloshing \citep{Murgiaetal11,Giacintuccietal11}. We therefore consider it likely that the position of the two lobes has been affected by sloshing, and that the cavity timescales are probably underestimated for the east lobe, and overestimated  for its western counterpart.

If the radiative age is an accurate estimate of the true age of the source, this suggests that the radio lobes have risen buoyantly to their current location, with only a limited period of supersonic jet expansion. If the jets are extended along the line of sight the lobes may be more distant from the core than they appear. However, the agreement between the various timescales suggests that any line-of-sight extension is probably small. As argued previously, the morphology of the radio source makes significant extension along the line of sight seem unlikely, particularly for the east lobe, which appears to be viewed edge-on.

Alternatively, the radio lobes might have remained close to the core because they have become approximately neutrally buoyant. There are a number of examples of group-central radio galaxies with steep spectral indices whose lobes nonetheless remain close to the group core, rather than having risen to large radii \citep[e.g., NGC~507, NGC~1407, NGC~5903,][]{Murgiaetal11,Giacintuccietal12,OSullivanetal18}. For radio galaxies whose jets have shut down, the buoyant rise of the cavities would be slowed if the filling factor of the relativistic plasma is reduced, for example if the radio lobes begin to fragment, and become interspersed with clouds or filaments of thermal IGM plasma \citep[see, e.g., discussion for the FR-I plumes and lobes of AWM~4,][]{OSullivanetal10}. In NGC~1550 we do see evidence of cavities at least partly coincident with the radio lobes, showing that they have excluded the IGM from a significant part of their volume, but we also see some IGM structures along lines of sight through the lobes. This could indicate that the volume occupied by the radio lobes is smaller than we have assumed. If the lobes are in fact complex structures with filaments or clouds of IGM gas penetrating them, this might also increase the drag, slowing any buoyant rise, and making it easier for sloshing motions to move the lobes.

Overall, the development of the radio source is likely to have been straightforward: 1) an initial phase of supersonic jet expansion, unaffected by any motions in the IGM, since the jet internal pressure would greatly exceed IGM pressure and the jet surface area would be small; 2) the inflation of the lobes, with slower, perhaps buoyancy-driven, expansion away from the core, during which IGM motions would have begun to affect the development of the source; 3) jet shutdown, after which the lobes would have expanded toward pressure equilibrium, while their position would have been determined by a combination of buoyant forces and IGM motion. It is in this last phase that the source likely became so strikingly asymmetric, and perhaps the west jet became kinked.

The similarity between the sloshing timescale and the radio source age raises the possibility that the AGN outburst was triggered by the same encounter which caused the sloshing. The impact of the sloshing motions would be limited in the galaxy core, but could potentially have precipitated gas into the central SMBH. If the AGN was triggered by cooling from the IGM, we might expect to see evidence of cooling filaments, ionized or molecular gas. We do not see such filaments in the \chandra\ data, and the galaxy is undetected in H$\alpha$ \citep{Lakhchauraetal18} and CO \citep{OSullivanetal18}. It is of course possible that the outburst exhausted the available reservoir of fuel, and (in combination with the sloshing) prevented further cooling. However, the absence of a large reservoir might indicate that only a limited fuel supply had built up before the sloshing motions disrupted it, initiating accretion and the AGN outburst. A more detailed understanding of the current state of the AGN and the morphology of the radio jets and lobes might give a clearer idea of its history and development.

\subsection{Identifying a potential perturber}

Sloshing is typically induced by a minor off-axis merger or fly-by gravitational interaction. We might therefore expect some evidence of a subhalo passing through or merging with the NGC~1550 group, perhaps a single large galaxy or smaller group of galaxies. However, NGC~1550 has relatively few massive galaxies in its neighbourhood.

Given the presence of a sloshing front to the east of the dominant elliptical, we expect the sloshing to be occuring in a plane east and west of the galaxy and along the line of sight. We can also expect any perturber to be close to this plane; it is less likely to be to the north or south of NGC~1550. Examining the volume of the group, we find that the second brightest galaxy, UGC~2998 is located $\sim$50\arcm\ WNW of NGC~1550, with a  velocity difference $\sim$400\kmps\ between the two. At our adopted distance for the group of 53~Mpc, this is equivalent to 770~kpc. The velocity dispersion of the group has been estimated to be $\sigma$=263$\pm$34\kmps \citep{Zhangetal17} or $\sim$300\kmps \citep{Sunetal03}. The velocity difference suggests that UGC~2998 is likely moving faster than this, but even assuming a velocity of 600\kmps, it would take $>$1~Gyr to travel from the group core to its current position. Our estimate of 40-80~Myr for the sloshing timescale only applies to the front we observe, and it is possible that the sloshing extends to larger radii, with older fronts still to be discovered. Nonetheless, UGC~2998 seems too distant to be a credible perturber. 

The next nearest galaxy of comparable brightness to NGC~1550 (1.6 mag fainter in $B$-band) is UGC~3008, $\sim$17\arcm\ (260~kpc) WNW of NGC~1550 and separated from it by $\sim$500\kmps. Again assuming a plane-of-sky velocity of 600\kmps, it would take $\sim$420~Myr to travel from the group core to its current position. If UGC~3008 is the perturber, this would certainly suggest that at least one more sloshing front should be present. Examination of the archival \xmm\ data does not reveal such a structure, but the effect of the point-spread function and instrumental structures may impact our ability to detect it. 

A third possibility is the elliptical IC~366, which lies $\sim$3\arcm\ SSW of NGC~1550. The two galaxies are separated by $<$25\kmps\ in velocity, suggesting that the motion of IC~366 relative to NGC~1550 is primarily in the plane of the sky. If IC~366 is moving away from NGC~1550 at 600\kmps, it would take $\sim$75~Myr to reach its current position, comparable to the longer sloshing timescale, and about a factor of 2 longer than the dynamical and radiative age estimates for the radio source. However, IC~366 is relatively small (2.3~mag fainter in $K$-band), would probably have to have passed very close to NGC~1550 to cause the sloshing, and its position to the south of NGC~1550 is not what we would expect given the shape of the sloshing front. Considering these different options, we conclude that we cannot be sure what has perturbed the NGC~1550 group, but that further investigation is warranted.


\subsection{Pressure balance in the lobes}
Our observations suggest that the radio lobes are old and probably no longer powered by the AGN. It is therefore reasonable to assume that they are passively evolving in pressure equilibrium with their surroundings. A comparison between the IGM thermal pressure and the apparent pressure in the lobes can provide information on the particle content of the lobes. Typically, the lobes of FR~I radio galaxies have apparent internal pressures well below that of their environment, suggesting that they are either out of equipartition, or that they contain a significant population of non-radiating particles \citep[e.g.,][]{Ferettietal92}.

We estimate the pressure inside the lobes $P_{\rm rad}$ based on the minimum energy magnetic field $B_{\rm min}$, their volume $V$ and the energy of the electron population $E_e$ using the equation:

\begin{equation}
{ P_{\rm rad}} = \frac{{ B_{\rm min}}^2}{2\mu_0\epsilon}+\frac{(1+k){ E_e}}{3 V\phi},
\end{equation}

\noindent where $\mu_0$ is the permeability of free space (4$\pi\times10^{-7}$ in SI units or 4$\pi$ in cgs units), $\phi$ is the filling factor of the relativistic radio-emitting plasma in the lobes, and k is the ratio of the energy in non-radiating particles to the energy in electrons. The $\epsilon$ factor represents the ordering of the magnetic field, with $\epsilon$=1 for a uniform field and $\epsilon$=3 for a tangled field. We initially adopt $\epsilon$=3, $\phi$=1 and k=0.

\begin{table}
\caption{Internal pressure ($P_{\rm rad}$) estimates for the east and west lobes, and ratios of external to internal pressure.}
\label{tab:Prad}
\centering
\begin{tabular}{lccc}
\hline
Lobe & $P_{\rm rad}$ & $P_{\rm IGM}$/$P_{\rm rad}$ & $(1+k)/\phi$ \\
 & (10$^{-12}$ erg cm$^{-3}$) & & \\
\hline
East & 5.20 & 12.7 & 340 \\
West & 4.24 & 9.4 & 160 \\
\hline
\end{tabular}
\end{table}

The resulting lobe pressures are shown in Table~\ref{tab:Prad}. As is usual for old FR~I radio lobes, the apparent internal pressure is significantly lower than the pressure of the surrounding IGM. This could indicate that the lobes are out of equipartition, that the filling factor of the radio plasma is less than unity, that there is a significant population of non-radiating particles in the lobes (e.g., entrained thermal plasma), and/or that our assumptions about the physical state of the plasma are incorrect. Taking into account the dependencies of $B_{\rm min}$ and $E_e$ on $k$ and $\phi$ we estimate the value of $(1+k)/\phi$ required to reach pressure balance (also shown in Table~\ref{tab:Prad}).

Given the visible cavities, it seems unlikely that a reduced filling factor can be responsible for the pressure imbalance, though it may contribute. An inverse-Compton detection would allow us to constrain the magnetic field and thus departure from equipartition, but we see no evidence of such emission. We are thus left to consider the possibility of additional non-radiating particles. These would need to be entrained and heated by the jets, to a temperature where their X-ray emission in the \chandra\ band was negligible, but not to the point where they would contribute to radio emission. We estimate the total energy of such a population would need to be $\sim$1.2$\times$10$^{57}$~erg, roughly 25-35~per~cent of the enthalpy of the radio lobes. Such a population of non-radiating particles is therefore at least possible, though our data do not allow us to meaningfully constrain their properties.

\section{Conclusions}
In this paper we have presented a detailed morphological and spectral analysis of the interesting radio source associated with NGC~1550. Multi-wavelength GMRT and archival $Chandra$ observations in combination with radio data from the literature, support the idea that a sloshing event has played a role in affecting the structure of the radio source in NGC~1550. A summary of our results is as follows.



\begin{itemize}
 
    \item Our GMRT radio images at 235 and 610~MHz show an asymmetric jet$-$lobe structure with a total size of $\sim$33~kpc at 235~MHz. The west jet is found to extend twice as far from the nucleus as the east jet, and the high resolution 610~MHz image reveals a kink to the south at the base of the west jet and on the other side a bending of the east jet towards the south while enclosed by the east lobe.
    
   \item The 235 $-$ 610~MHz spectral index map of the source reveals that both radio lobes present steep spectral index values ranging between about $-1.2$ and $-1.8$ indicating the presence of an old electron population. The eastern jet which is seen to be enclosed by the eastern lobe, and the more extended and visible western jet, present a similar spectral index value of $\alpha_{235 MHz}^{610 MHz}\approx-1.2$, which indicates past activity.
   
    
    \item Combining the analysed GMRT data with radio flux density  measurements from the literature between 88 and 2380~MHz, we find that a simple power-law model fit for the radio source gives a steep spectrum of $\alpha_{\mathrm{inj}}=-1.37_{-0.06}^{+0.01}$, indicating that it is dominated by the radio lobes. 
    
      \item The radio structure is found to be correlated with the X-ray emission. The brightest X-ray emission, similar to the radio one, is asymmetric being more extended to the west. We observe potential cavities, rims and arms of gas associated with both radio lobes which may have been uplifted by the cavity expansion. We estimate the jet power to be a few 10$^{42}$~erg~s$^{-1}$, which is comparable to the bolometric X-ray luminosity within the region affected by the jets, [1.81$\pm$0.02]$\times$10$^{42}$~erg~s$^{-1}$. This  suggests that the AGN is capable of balancing radiative losses in the cooling region. A much more powerful outburst would be required to drive gas out of the group.

    
    
    \item We find a minimum energy magnetic field of $B_{\mathrm{min}}\approx11 \mathrm{\mu G}$ for the source in total. Assuming a magnetic field strength which minimizes the combined energy in the magnetic field and the relativistic particles, and a lower bound on the electron energy distribution of $\gamma_{\rm min}=100$ we fit a JP model in the observed 235 $-$ 610~MHz spectral index trend along the source and find a total radiative age of $t_{\rm rad} \approx $ 33~Myr for the radio source in NGC~1550. The radiative ages in the various components of the radio source range between $t_{\rm rad}\approx $ 20 $-$ 36~Myr. 
    
    
\item We find indications of sloshing in NGC~1550 from the observed arc-shaped structure in the X-ray residual map. We find that the overall radiative age estimate derived from the radio observations ($\sim$33~Myr) is in reasonable agreement with the range of dynamical timescale estimates derived from the X-ray data (20 $-$ 50 Myr), and is somewhat shorter than the shortest timescale estimate for sloshing motions (40 $-$ 80 Myr). The radiative age estimates for the individual lobes ($\sim$22 $-$ 25~Myr) are at the lower end of the range of dynamical age estimates, and about half the sloshing timescale. From our data we cannot determine with certainty which galaxy is the perturber of NGC~1550 with further investigation warranted.

\item We conclude that the observed radio/X-ray structure in NGC~1550 is most likely the combination of sloshing motions with effects of buoyant forces from jet/lobe growth in the IGM. Although NGC~1550 has been considered a relaxed system hosting a decaying radio source, we find evidence that it has undergone a recent minor merger. \\

Overall, the combined radio/X-ray study of NGC~1550 presented in this paper reveals our ability to assess the dynamical state and earlier merging history of the group. We suggest that the source underwent an initial short phase of supersonic jet expansion followed by the inflation of the lobes away from the core during which IGM motions would begin to affect the source development. In the final phase, the jet turned off leaving the lobes to expand towards pressure equilibrium, with their structure governed by buoyant forces and IGM motion. It is most likely that in this final phase the asymmetric and kinked morphology of the radio source arose. As the sloshing timescale and the radio source age are found to be similar it is possible that the AGN outburst was triggered by the same encounter which caused the sloshing. In the future, this work will benefit from high frequency radio observations of the NGC~1550 radio source that will help constrain the radiative age estimate and from better X-ray data that will provide more information regarding the existence of another sloshing front further out, which could possibly constrain the nature of the perturber and provide clearer insight into the relation between the processes involved.

\end{itemize}

\section*{Acknowledgements}
K. Kolokythas is supported by the Centre of Space Research at North-West University. E. O'Sullivan acknowledges support through \chandra\ award number GO7-18126X issued by the \chandra\ X-ray Center, which is operated by the Smithsonian Astrophysical Observatory for and on behalf of the National Aeronautics Space Administration under contract NAS8-03060. S. Giacintucci acknowledges support for basic research in radio astronomy at the Naval Research Laboratory by 6.1 Base funding. I. Loubser is funded by the National Research Foundation (NRF) of South Africa. GMRT is a national facility operated by the National Center for Radio Astrophysics (NCRA) of the Tata Institute for Fundamental Research (TIFR). This research has made use of the NASA/IPAC Extragalactic Database (NED) which is operated by the Jet Propulsion Laboratory, California Institute of Technology, under contract with the National Aeronautics and Space Administration. 



\bibliographystyle{mnras}
\bibliography{paper.bib} 

\begin{thebibliography}{}
\makeatletter
\relax
\def\mn@urlcharsother{\let\do\@makeother \do\$\do\&\do\#\do\^\do\_\do\%\do\~}
\def\mn@doi{\begingroup\mn@urlcharsother \@ifnextchar [ {\mn@doi@}
  {\mn@doi@[]}}
\def\mn@doi@[#1]#2{\def\@tempa{#1}\ifx\@tempa\@empty \href
  {http://dx.doi.org/#2} {doi:#2}\else \href {http://dx.doi.org/#2} {#1}\fi
  \endgroup}
\def\mn@eprint#1#2{\mn@eprint@#1:#2::\@nil}
\def\mn@eprint@arXiv#1{\href {http://arxiv.org/abs/#1} {{\tt arXiv:#1}}}
\def\mn@eprint@dblp#1{\href {http://dblp.uni-trier.de/rec/bibtex/#1.xml}
  {dblp:#1}}
\def\mn@eprint@#1:#2:#3:#4\@nil{\def\@tempa {#1}\def\@tempb {#2}\def\@tempc
  {#3}\ifx \@tempc \@empty \let \@tempc \@tempb \let \@tempb \@tempa \fi \ifx
  \@tempb \@empty \def\@tempb {arXiv}\fi \@ifundefined
  {mn@eprint@\@tempb}{\@tempb:\@tempc}{\expandafter \expandafter \csname
  mn@eprint@\@tempb\endcsname \expandafter{\@tempc}}}

\bibitem[\protect\citeauthoryear{{Arnaud}}{{Arnaud}}{1996}]{Arnaud96}
{Arnaud} K.~A.,  1996, in {Jacoby} G.~H.,  {Barnes} J.,  eds,  Astronomical
  Society of the Pacific Conference Series Vol. 101, Astronomical Data Analysis
  Software and Systems V. p.~17

\bibitem[\protect\citeauthoryear{{Ascasibar} \& {Markevitch}}{{Ascasibar} \&
  {Markevitch}}{2006}]{AscasibarMarkevitch06}
{Ascasibar} Y.,  {Markevitch} M.,  2006, \mn@doi [ApJ] {10.1086/506508}, \href
  {http://adsabs.harvard.edu/abs/2006ApJ...650..102A} {650, 102}

\bibitem[\protect\citeauthoryear{{Baars}, {Genzel}, {Pauliny-Toth}  \&
  {Witzel}}{{Baars} et~al.}{1977}]{Baarsetal77}
{Baars} J.~W.~M.,  {Genzel} R.,  {Pauliny-Toth} I.~I.~K.,   {Witzel} A.,  1977,
  A\&A, \href {http://adsabs.harvard.edu/abs/1977A\%26A....61...99B} {61, 99}

\bibitem[\protect\citeauthoryear{{Balbus} \& {Soker}}{{Balbus} \&
  {Soker}}{1990}]{BalbusSoker90}
{Balbus} S.~A.,  {Soker} N.,  1990, \mn@doi [\apj] {10.1086/168926}, \href
  {https://ui-adsabs-harvard-edu.ezp-prod1.hul.harvard.edu/abs/1990ApJ...357..353B}
  {357, 353}

\bibitem[\protect\citeauthoryear{{B{\^i}rzan}, {Rafferty}, {McNamara}, {Wise}
  \& {Nulsen}}{{B{\^i}rzan} et~al.}{2004}]{Birzanetal04}
{B{\^i}rzan} L.,  {Rafferty} D.~A.,  {McNamara} B.~R.,  {Wise} M.~W.,
  {Nulsen} P.~E.~J.,  2004, \mn@doi [ApJ] {10.1086/383519}, \href
  {http://adsabs.harvard.edu/cgi-bin/nph-bib_query?bibcode=2004ApJ...607..800B&db_key=AST}
  {607, 800}

\bibitem[\protect\citeauthoryear{{Brown}, {Jannuzi}, {Floyd}  \&
  {Mould}}{{Brown} et~al.}{2011}]{Brownetal11}
{Brown} M.~J.~I.,  {Jannuzi} B.~T.,  {Floyd} D.~J.~E.,   {Mould} J.~R.,  2011,
  \mn@doi [ApJ] {10.1088/2041-8205/731/2/L41}, \href
  {http://adsabs.harvard.edu/abs/2011ApJ...731L..41B} {731, L41}

\bibitem[\protect\citeauthoryear{{Bykov}, {Churazov}, {Ferrari}, {Forman},
  {Kaastra}, {Klein}, {Markevitch}  \& {de Plaa}}{{Bykov}
  et~al.}{2015}]{Bykov15}
{Bykov} A.~M.,  {Churazov} E.~M.,  {Ferrari} C.,  {Forman} W.~R.,  {Kaastra}
  J.~S.,  {Klein} U.,  {Markevitch} M.,   {de Plaa} J.,  2015, \mn@doi [\ssr]
  {10.1007/s11214-014-0129-4}, \href
  {https://ui.adsabs.harvard.edu/abs/2015SSRv..188..141B} {188, 141}

\bibitem[\protect\citeauthoryear{{Calzadilla} et~al.,}{{Calzadilla}
  et~al.}{2019}]{Calzadillaetal19}
{Calzadilla} M.~S.,  et~al., 2019, ApJ, 875, 65

\bibitem[\protect\citeauthoryear{{Canning} et~al.,}{{Canning}
  et~al.}{2017}]{Canningetal17}
{Canning} R.~E.~A.,  et~al., 2017, MNRAS, 464, 2896

\bibitem[\protect\citeauthoryear{{Chandra}, {Ray}  \& {Bhatnagar}}{{Chandra}
  et~al.}{2004}]{Chandra04}
{Chandra} P.,  {Ray} A.,   {Bhatnagar} S.,  2004, \mn@doi [\apj]
  {10.1086/422675}, \href {http://cdsads.u-strasbg.fr/abs/2004ApJ...612..974C}
  {612, 974}

\bibitem[\protect\citeauthoryear{{Churazov}, {Forman}, {Jones}  \&
  {B{\"o}hringer}}{{Churazov} et~al.}{2003}]{Churazovetal03}
{Churazov} E.,  {Forman} W.,  {Jones} C.,   {B{\"o}hringer} H.,  2003, \mn@doi
  [\apj] {10.1086/374923}, \href
  {https://ui-adsabs-harvard-edu.ezp-prod1.hul.harvard.edu/abs/2003ApJ...590..225C}
  {590, 225}

\bibitem[\protect\citeauthoryear{{David} et~al.,}{{David}
  et~al.}{2011}]{Davidetal11}
{David} L.~P.,  et~al., 2011, \mn@doi [ApJ] {10.1088/0004-637X/728/2/162},
  \href {http://adsabs.harvard.edu/abs/2011ApJ...728..162D} {728, 162}

\bibitem[\protect\citeauthoryear{{Davies} \& {Birkinshaw}}{{Davies} \&
  {Birkinshaw}}{1986}]{DaviesBirkinshaw86}
{Davies} R.~L.,  {Birkinshaw} M.,  1986, \mn@doi [ApJ] {10.1086/184650}, \href
  {http://adsabs.harvard.edu/abs/1986ApJ...303L..45D} {303, L45}

\bibitem[\protect\citeauthoryear{{Dressel} \& {Condon}}{{Dressel} \&
  {Condon}}{1978}]{DresselCondon78}
{Dressel} L.~L.,  {Condon} J.~J.,  1978, \mn@doi [ApJS] {10.1086/190491}, \href
  {http://adsabs.harvard.edu/abs/1978ApJS...36...53D} {36, 53}

\bibitem[\protect\citeauthoryear{{Dunn}, {Allen}, {Taylor}, {Shurkin},
  {Gentile}, {Fabian}  \& {Reynolds}}{{Dunn} et~al.}{2010}]{Dunnetal10}
{Dunn} R.~J.~H.,  {Allen} S.~W.,  {Taylor} G.~B.,  {Shurkin} K.~F.,  {Gentile}
  G.,  {Fabian} A.~C.,   {Reynolds} C.~S.,  2010, \mn@doi [MNRAS]
  {10.1111/j.1365-2966.2010.16314.x}, \href
  {http://adsabs.harvard.edu/abs/2010MNRAS.404..180D} {404, 180}

\bibitem[\protect\citeauthoryear{{Dupke}, {White}  \& {Bregman}}{{Dupke}
  et~al.}{2007}]{Dupkeetal07}
{Dupke} R.,  {White} Raymond~E. I.,   {Bregman} J.~N.,  2007, \mn@doi [\apj]
  {10.1086/522194}, \href
  {https://ui.adsabs.harvard.edu/abs/2007ApJ...671..181D} {671, 181}

\bibitem[\protect\citeauthoryear{{Ehlert}, {McDonald}, {David}, {Miller}  \&
  {Bautz}}{{Ehlert} et~al.}{2015}]{Ehlertetal14}
{Ehlert} S.,  {McDonald} M.,  {David} L.~P.,  {Miller} E.~D.,   {Bautz} M.~W.,
  2015, \mn@doi [ApJ] {10.1088/0004-637X/799/2/174}, 799, 174

\bibitem[\protect\citeauthoryear{{Fabian}, {Crawford}, {Edge}  \&
  {Mushotzky}}{{Fabian} et~al.}{1994}]{Fabianetal94}
{Fabian} A.~C.,  {Crawford} C.~S.,  {Edge} A.~C.,   {Mushotzky} R.~F.,  1994,
  MNRAS, 267, 779

\bibitem[\protect\citeauthoryear{{Fabian} et~al.,}{{Fabian}
  et~al.}{2011}]{Fabianetal11}
{Fabian} A.~C.,  et~al., 2011, \mn@doi [MNRAS]
  {10.1111/j.1365-2966.2011.19402.x}, \href
  {http://adsabs.harvard.edu/abs/2011MNRAS.418.2154F} {418, 2154}

\bibitem[\protect\citeauthoryear{{Feretti}, {Perola}  \& {Fanti}}{{Feretti}
  et~al.}{1992}]{Ferettietal92}
{Feretti} L.,  {Perola} G.~C.,   {Fanti} R.,  1992, A\&A, \href
  {http://adsabs.harvard.edu/abs/1992A\%26A...265....9F} {265, 9}

\bibitem[\protect\citeauthoryear{{Freeman}, {Doe}  \&
  {Siemiginowska}}{{Freeman} et~al.}{2001}]{Freemanetal01}
{Freeman} P.,  {Doe} S.,   {Siemiginowska} A.,  2001, in {J.-L.~Starck \&
  F.~D.~Murtagh} ed.,  Society of Photo-Optical Instrumentation Engineers
  (SPIE) Conference Series Vol. 4477, Society of Photo-Optical Instrumentation
  Engineers (SPIE) Conference Series. p.~76 (\mn@eprint {}
  {arXiv:astro-ph/0108426}), \mn@doi{10.1117/12.447161}

\bibitem[\protect\citeauthoryear{{Gastaldello}, {Buote}, {Humphrey},
  {Zappacosta}, {Bullock}, {Brighenti}  \& {Mathews}}{{Gastaldello}
  et~al.}{2007}]{Gastaldelloetal07}
{Gastaldello} F.,  {Buote} D.~A.,  {Humphrey} P.~J.,  {Zappacosta} L.,
  {Bullock} J.~S.,  {Brighenti} F.,   {Mathews} W.~G.,  2007, \mn@doi [ApJ]
  {10.1086/521519}, \href {http://adsabs.harvard.edu/abs/2007ApJ...669..158G}
  {669, 158}

\bibitem[\protect\citeauthoryear{{Gastaldello}, {Buote}, {Temi}, {Brighenti},
  {Mathews}  \& {Ettori}}{{Gastaldello} et~al.}{2009}]{Gastaldelloetal09}
{Gastaldello} F.,  {Buote} D.~A.,  {Temi} P.,  {Brighenti} F.,  {Mathews}
  W.~G.,   {Ettori} S.,  2009, \mn@doi [ApJ] {10.1088/0004-637X/693/1/43},
  \href {http://adsabs.harvard.edu/abs/2009ApJ...693...43G} {693, 43}

\bibitem[\protect\citeauthoryear{{Gastaldello} et~al.,}{{Gastaldello}
  et~al.}{2013}]{Gastaldelloetal13}
{Gastaldello} F.,  et~al., 2013, \mn@doi [ApJ] {10.1088/0004-637X/770/1/56},
  \href {http://adsabs.harvard.edu/abs/2013ApJ...770...56G} {770, 56}

\bibitem[\protect\citeauthoryear{{Ghizzardi}, {De Grandi}  \&
  {Molendi}}{{Ghizzardi} et~al.}{2013}]{Ghizzardietal13}
{Ghizzardi} S.,  {De Grandi} S.,   {Molendi} S.,  2013, \mn@doi
  [Astron.~Nachr.] {10.1002/asna.201211871}, \href
  {http://adsabs.harvard.edu/abs/2013AN....334..422G} {334, 422}

\bibitem[\protect\citeauthoryear{{Giacintucci} et~al.,}{{Giacintucci}
  et~al.}{2008}]{Giacintuccietal08}
{Giacintucci} S.,  et~al., 2008, \mn@doi [ApJ] {10.1086/589280}, \href
  {http://adsabs.harvard.edu/abs/2008ApJ...682..186G} {682, 186}

\bibitem[\protect\citeauthoryear{{Giacintucci} et~al.,}{{Giacintucci}
  et~al.}{2011}]{Giacintuccietal11}
{Giacintucci} S.,  et~al., 2011, \mn@doi [ApJ] {10.1088/0004-637X/732/2/95},
  \href {http://adsabs.harvard.edu/abs/2011ApJ...732...95G} {732, 95}

\bibitem[\protect\citeauthoryear{{Giacintucci} et~al.,}{{Giacintucci}
  et~al.}{2012}]{Giacintuccietal12}
{Giacintucci} S.,  et~al., 2012, \mn@doi [ApJ] {10.1088/0004-637X/755/2/172},
  \href {http://adsabs.harvard.edu/abs/2012ApJ...755..172G} {755, 172}

\bibitem[\protect\citeauthoryear{{Gitti}, {McNamara}, {Nulsen}  \&
  {Wise}}{{Gitti} et~al.}{2007}]{Gittietal07}
{Gitti} M.,  {McNamara} B.~R.,  {Nulsen} P.~E.~J.,   {Wise} M.~W.,  2007,
  \mn@doi [ApJ] {10.1086/512800}, \href
  {http://adsabs.harvard.edu/abs/2007ApJ...660.1118G} {660, 1118}

\bibitem[\protect\citeauthoryear{{Grevesse} \& {Sauval}}{{Grevesse} \&
  {Sauval}}{1998}]{GrevesseSauval98}
{Grevesse} N.,  {Sauval} A.~J.,  1998, Space Sci.~Rev., \href
  {http://adsabs.harvard.edu/cgi-bin/nph-bib_query?bibcode=1998SSRv...85..161G&db_key=AST}
  {85, 161}

\bibitem[\protect\citeauthoryear{{Hardcastle}, {Birkinshaw}  \&
  {Worrall}}{{Hardcastle} et~al.}{2001}]{HardcastleBW01}
{Hardcastle} M.~J.,  {Birkinshaw} M.,   {Worrall} D.~M.,  2001, \mn@doi
  [\mnras] {10.1111/j.1365-2966.2001.04699.x}, \href
  {https://ui.adsabs.harvard.edu/abs/2001MNRAS.326.1499H} {326, 1499}

\bibitem[\protect\citeauthoryear{{Hashimoto} \& {Oemler}}{{Hashimoto} \&
  {Oemler}}{2000}]{HashimotoOemler00}
{Hashimoto} Y.,  {Oemler} Augustus J.,  2000, \mn@doi [\apj] {10.1086/308383},
  \href {https://ui.adsabs.harvard.edu/abs/2000ApJ...530..652H} {530, 652}

\bibitem[\protect\citeauthoryear{Hurley-Walker et~al.,}{Hurley-Walker
  et~al.}{2016}]{HurleyWalkeretal17}
Hurley-Walker N.,  et~al., 2016, \mn@doi [MNRAS] {10.1093/mnras/stw2337}, 464,
  1146

\bibitem[\protect\citeauthoryear{{Intema}}{{Intema}}{2014}]{Intema14}
{Intema} H.~T.,  2014, in Astronomical Society of India Conference Series.
  (\mn@eprint {arXiv} {1402.4889})

\bibitem[\protect\citeauthoryear{{Intema}, {van der Tol}, {Cotton}, {Cohen},
  {van Bemmel}  \& {R{\"o}ttgering}}{{Intema} et~al.}{2009}]{Intemaetal09}
{Intema} H.~T.,  {van der Tol} S.,  {Cotton} W.~D.,  {Cohen} A.~S.,  {van
  Bemmel} I.~M.,   {R{\"o}ttgering} H.~J.~A.,  2009, \mn@doi [A\&A]
  {10.1051/0004-6361/200811094}, 501, 1185

\bibitem[\protect\citeauthoryear{{Intema}, {Jagannathan}, {Mooley}  \&
  {Frail}}{{Intema} et~al.}{2017}]{Intemaetal17}
{Intema} H.~T.,  {Jagannathan} P.,  {Mooley} K.~P.,   {Frail} D.~A.,  2017,
  A\&A, \href {http://adsabs.harvard.edu/abs/2017A\%26A...598A..78I} {598, 78}

\bibitem[\protect\citeauthoryear{{Jaffe} \& {Perola}}{{Jaffe} \&
  {Perola}}{1973}]{JaffePerola73}
{Jaffe} W.~J.,  {Perola} G.~C.,  1973, \aap, \href
  {https://ui.adsabs.harvard.edu/abs/1973A&A....26..423J} {26, 423}

\bibitem[\protect\citeauthoryear{{Jones}, {Ponman}, {Horton}, {Babul},
  {Ebeling}  \& {Burke}}{{Jones} et~al.}{2003}]{Jonesetal03}
{Jones} L.~R.,  {Ponman} T.~J.,  {Horton} A.,  {Babul} A.,  {Ebeling} H.,
  {Burke} D.~J.,  2003, MNRAS, \href
  {http://adsabs.harvard.edu/cgi-bin/nph-bib_query?bibcode=2003MNRAS.343..627J&db_key=AST}
  {343, 627}

\bibitem[\protect\citeauthoryear{{Kalberla}, {Burton}, {Hartmann}, {Arnal},
  {Bajaja}, {Morras}  \& {P{\"o}ppel}}{{Kalberla}
  et~al.}{2005}]{Kalberlaetal05}
{Kalberla} P.~M.~W.,  {Burton} W.~B.,  {Hartmann} D.,  {Arnal} E.~M.,  {Bajaja}
  E.,  {Morras} R.,   {P{\"o}ppel} W.~G.~L.,  2005, \mn@doi [A\&A] {10.1051
  /0004-6361:20041864}, \href
  {http://adsabs.harvard.edu/abs/2005A\%26A...440..775K} {440, 775}

\bibitem[\protect\citeauthoryear{{Kardashev}}{{Kardashev}}{1962}]{Kardashev62}
{Kardashev} N.~S.,  1962, Soviet Astronomy, \href
  {http://adsabs.harvard.edu/abs/1962SvA.....6..317K} {6, 317}

\bibitem[\protect\citeauthoryear{{Kirkpatrick}, {Gitti}, {Cavagnolo},
  {McNamara}, {David}, {Nulsen}  \& {Wise}}{{Kirkpatrick}
  et~al.}{2009}]{Kirkpatricketal09}
{Kirkpatrick} C.~C.,  {Gitti} M.,  {Cavagnolo} K.~W.,  {McNamara} B.~R.,
  {David} L.~P.,  {Nulsen} P.~E.~J.,   {Wise} M.~W.,  2009, \mn@doi [ApJ]
  {10.1088/0004-637X/707/1/L69}, \href
  {http://adsabs.harvard.edu/abs/2009ApJ...707L..69K} {707, L69}

\bibitem[\protect\citeauthoryear{{Kirkpatrick}, {McNamara}  \&
  {Cavagnolo}}{{Kirkpatrick} et~al.}{2011}]{Kirkpatricketal11}
{Kirkpatrick} C.~C.,  {McNamara} B.~R.,   {Cavagnolo} K.~W.,  2011, \mn@doi
  [ApJ] {10.1088/2041-8205/731/2/L23}, \href
  {http://adsabs.harvard.edu/abs/2011ApJ...731L..23K} {731, L23+}

\bibitem[\protect\citeauthoryear{{Kolokythas}, {O'Sullivan}, {Giacintucci},
  {Raychaudhury}, {Ishwara-Chandra}, {Worrall}  \& {Birkinshaw}}{{Kolokythas}
  et~al.}{2015}]{Kolokythasetal15}
{Kolokythas} K.,  {O'Sullivan} E.,  {Giacintucci} S.,  {Raychaudhury} S.,
  {Ishwara-Chandra} C.~H.,  {Worrall} D.~M.,   {Birkinshaw} M.,  2015, \mn@doi
  [MNRAS] {10.1093/mnras/stv665}, \href
  {http://adsabs.harvard.edu/abs/2015MNRAS.450.1732K} {450, 1732}

\bibitem[\protect\citeauthoryear{{Kolokythas}, {O'Sullivan}, {Raychaudhury},
  {Giacintucci}, {Gitti}  \& {Babul}}{{Kolokythas}
  et~al.}{2018}]{Kolokythasetal18}
{Kolokythas} K.,  {O'Sullivan} E.,  {Raychaudhury} S.,  {Giacintucci} S.,
  {Gitti} M.,   {Babul} A.,  2018, \mn@doi [MNRAS] {10.1093/mnras/sty2030},
  \href {http://adsabs.harvard.edu/abs/2018MNRAS.481.1550K} {481, 1550}

\bibitem[\protect\citeauthoryear{{Kolokythas}, {O'Sullivan}, {Intema},
  {Raychaudhury}, {Babul}, {Giacintucci}  \& {Gitti}}{{Kolokythas}
  et~al.}{2019}]{Kolokythasetal19}
{Kolokythas} K.,  {O'Sullivan} E.,  {Intema} H.,  {Raychaudhury} S.,  {Babul}
  A.,  {Giacintucci} S.,   {Gitti} M.,  2019, \mn@doi [\mnras]
  {10.1093/mnras/stz2082}, \href
  {https://ui.adsabs.harvard.edu/abs/2019MNRAS.489.2488K} {489, 2488}

\bibitem[\protect\citeauthoryear{{Lakhchaura} et~al.,}{{Lakhchaura}
  et~al.}{2018}]{Lakhchauraetal18}
{Lakhchaura} K.,  et~al., 2018, \mn@doi [MNRAS] {10.1093/mnras/sty2565}, 481,
  4472

\bibitem[\protect\citeauthoryear{{Liu} et~al.,}{{Liu}
  et~al.}{2019}]{Wenhaoetal19}
{Liu} W.,  et~al., 2019, \mn@doi [\mnras] {10.1093/mnras/stz229}, \href
  {https://ui.adsabs.harvard.edu/abs/2019MNRAS.484.3376L} {484, 3376}

\bibitem[\protect\citeauthoryear{{Machacek}, {Jerius}, {Kraft}, {Forman},
  {Jones}, {Randall}, {Giancintucci}  \& {Sun}}{{Machacek}
  et~al.}{2011}]{Machaceketal11}
{Machacek} M.~E.,  {Jerius} D.,  {Kraft} R.~P.,  {Forman} W.~R.,  {Jones} C.,
  {Randall} S.,  {Giancintucci} S.,   {Sun} M.,  2011, ApJ, \href
  {http://adsabs.harvard.edu/abs/2011arXiv1108.5229M} {743, 15}

\bibitem[\protect\citeauthoryear{{Mamon}}{{Mamon}}{2000}]{Mamon00}
{Mamon} G.~A.,  2000, in {Combes} F.,  {Mamon} G.~A.,   {Charmandaris} V.,
  eds,  Astronomical Society of the Pacific Conference Series Vol. 197,
  Dynamics of Galaxies: from the Early Universe to the Present. p.~377
  (\mn@eprint {arXiv} {astro-ph/9911333})

\bibitem[\protect\citeauthoryear{{Markevitch} \& {Vikhlinin}}{{Markevitch} \&
  {Vikhlinin}}{2007}]{MarkevitchVikhlinin07}
{Markevitch} M.,  {Vikhlinin} A.,  2007, \mn@doi [Phys. Rep.]
  {10.1016/j.physrep.2007.01.001}, \href
  {http://adsabs.harvard.edu/abs/2007PhR...443....1M} {443, 1}

\bibitem[\protect\citeauthoryear{{Markevitch}, {Vikhlinin}  \&
  {Mazzotta}}{{Markevitch} et~al.}{2001}]{Markevitchetal01}
{Markevitch} M.,  {Vikhlinin} A.,   {Mazzotta} P.,  2001, ApJ, 562, L153

\bibitem[\protect\citeauthoryear{{Mazzotta}, {Markevitch}, {Vikhlinin},
  {Forman}, {David}  \& {van Speybroeck}}{{Mazzotta}
  et~al.}{2001}]{Mazzottaetal01}
{Mazzotta} P.,  {Markevitch} M.,  {Vikhlinin} A.,  {Forman} W.~R.,  {David}
  L.~P.,   {van Speybroeck} L.,  2001, \mn@doi [\apj] {10.1086/321484}, \href
  {https://ui.adsabs.harvard.edu/abs/2001ApJ...555..205M} {555, 205}

\bibitem[\protect\citeauthoryear{{McNamara} \& {Nulsen}}{{McNamara} \&
  {Nulsen}}{2007}]{McNamaraNulsen07}
{McNamara} B.~R.,  {Nulsen} P.~E.~J.,  2007, \mn@doi [ARA\&A]
  {10.1146/annurev.astro.45.051806.110625}, \href
  {http://adsabs.harvard.edu/abs/2007ARA\%26A..45..117M} {45, 117}

\bibitem[\protect\citeauthoryear{{McNamara} \& {Nulsen}}{{McNamara} \&
  {Nulsen}}{2012}]{McNamaraNulsen12}
{McNamara} B.~R.,  {Nulsen} P.~E.~J.,  2012, \mn@doi [New Journal of Physics]
  {10.1088/1367-2630/14/5/055023}, \href
  {http://adsabs.harvard.edu/abs/2012NJPh...14e5023M} {14, 055023}

\bibitem[\protect\citeauthoryear{{McNamara}, {Russell}, {Nulsen}, {Hogan},
  {Fabian}, {Pulido}  \& {Edge}}{{McNamara} et~al.}{2016}]{McNamaraetal16}
{McNamara} B.~R.,  {Russell} H.~R.,  {Nulsen} P.~E.~J.,  {Hogan} M.~T.,
  {Fabian} A.~C.,  {Pulido} F.,   {Edge} A.~C.,  2016, \mn@doi [ApJ]
  {10.3847/0004-637X/830/2/79}, \href
  {http://adsabs.harvard.edu/abs/2016ApJ...830...79M} {830, 79}

\bibitem[\protect\citeauthoryear{{Mulchaey} \& {Zabludoff}}{{Mulchaey} \&
  {Zabludoff}}{1998}]{MulchaeyZabludoff98}
{Mulchaey} J.~S.,  {Zabludoff} A.~I.,  1998, \mn@doi [\apj] {10.1086/305356},
  \href {https://ui.adsabs.harvard.edu/abs/1998ApJ...496...73M} {496, 73}

\bibitem[\protect\citeauthoryear{{Murgia}}{{Murgia}}{2001}]{Murgia01}
{Murgia} M.,  2001, PhD thesis, Universit{\'a} di Bologna

\bibitem[\protect\citeauthoryear{{Murgia}}{{Murgia}}{2003}]{Murgia03}
{Murgia} M.,  2003, \mn@doi [\pasa] {10.1071/AS02033}, \href
  {https://ui.adsabs.harvard.edu/abs/2003PASA...20...19M} {20, 19}

\bibitem[\protect\citeauthoryear{{Murgia} et~al.,}{{Murgia}
  et~al.}{2011}]{Murgiaetal11}
{Murgia} M.,  et~al., 2011, \mn@doi [A\&A] {10.1051/0004-6361/201015302}, \href
  {http://adsabs.harvard.edu/abs/2011A\%26A...526A.148M} {526, A148+}

\bibitem[\protect\citeauthoryear{{Myers} \& {Spangler}}{{Myers} \&
  {Spangler}}{1985}]{MyersSpangler85}
{Myers} S.~T.,  {Spangler} S.~R.,  1985, \mn@doi [ApJ] {10.1086/163040}, \href
  {http://adsabs.harvard.edu/abs/1985ApJ...291...52M} {291, 52}

\bibitem[\protect\citeauthoryear{{O'Sullivan}, {Giacintucci}, {David},
  {Vrtilek}  \& {Raychaudhury}}{{O'Sullivan} et~al.}{2010}]{OSullivanetal10}
{O'Sullivan} E.,  {Giacintucci} S.,  {David} L.~P.,  {Vrtilek} J.~M.,
  {Raychaudhury} S.,  2010, \mn@doi [MNRAS] {10.1111/j.1365-2966.2010.16895.x},
  \href {http://adsabs.harvard.edu/abs/2010MNRAS.407..321O} {407, 321}

\bibitem[\protect\citeauthoryear{{O'Sullivan}, {Giacintucci}, {David},
  {Vrtilek}  \& {Raychaudhury}}{{O'Sullivan} et~al.}{2011a}]{OSullivanetal11a}
{O'Sullivan} E.,  {Giacintucci} S.,  {David} L.~P.,  {Vrtilek} J.~M.,
  {Raychaudhury} S.,  2011a, \mn@doi [MNRAS]
  {10.1111/j.1365-2966.2010.17812.x}, \href
  {http://adsabs.harvard.edu/abs/2011MNRAS.411.1833O} {411, 1833}

\bibitem[\protect\citeauthoryear{{O'Sullivan}, {Worrall}, {Birkinshaw},
  {Trinchieri}, {Wolter}, {Zezas}  \& {Giacintucci}}{{O'Sullivan}
  et~al.}{2011b}]{OSullivanetal11c}
{O'Sullivan} E.,  {Worrall} D.~M.,  {Birkinshaw} M.,  {Trinchieri} G.,
  {Wolter} A.,  {Zezas} A.,   {Giacintucci} S.,  2011b, \mn@doi [MNRAS]
  {10.1111/j.1365-2966.2011.19239.x}, \href
  {http://adsabs.harvard.edu/abs/2011MNRAS.tmp.1180O} {416, 2916}

\bibitem[\protect\citeauthoryear{{O'Sullivan}, {Giacintucci}, {David}, {Gitti},
  {Vrtilek}, {Raychaudhury}  \& {Ponman}}{{O'Sullivan}
  et~al.}{2011c}]{OSullivanetal11b}
{O'Sullivan} E.,  {Giacintucci} S.,  {David} L.~P.,  {Gitti} M.,  {Vrtilek}
  J.~M.,  {Raychaudhury} S.,   {Ponman} T.~J.,  2011c, \mn@doi [ApJ]
  {10.1088/0004-637X/735/1/11}, \href
  {http://adsabs.harvard.edu/abs/2011ApJ...735...11O} {735, 11}

\bibitem[\protect\citeauthoryear{{O'Sullivan}, {David}  \&
  {Vrtilek}}{{O'Sullivan} et~al.}{2014}]{OSullivanetal14a}
{O'Sullivan} E.,  {David} L.~P.,   {Vrtilek} J.~M.,  2014, \mn@doi [MNRAS]
  {10.1093/mnras/stt1926}, \href
  {http://adsabs.harvard.edu/abs/2014MNRAS.437..730O} {437, 730}

\bibitem[\protect\citeauthoryear{{O'Sullivan} et~al.,}{{O'Sullivan}
  et~al.}{2017}]{OSullivanetal17}
{O'Sullivan} E.,  et~al., 2017, \mn@doi [MNRAS] {10.1093/mnras/stx2078}, 472,
  1482

\bibitem[\protect\citeauthoryear{{O'Sullivan}, {Kolokythas}, {Kantharia},
  {Raychaudhury}, {David}  \& {Vrtilek}}{{O'Sullivan}
  et~al.}{2018}]{OSullivanetal18}
{O'Sullivan} E.,  {Kolokythas} K.,  {Kantharia} N.~G.,  {Raychaudhury} S.,
  {David} L.~P.,   {Vrtilek} J.~M.,  2018, \mn@doi [MNRAS]
  {10.1093/mnras/stx2702}, \href
  {http://adsabs.harvard.edu/abs/2018MNRAS.473.5248O} {473, 5248}

\bibitem[\protect\citeauthoryear{{Parma}, {Murgia}, {de Ruiter}, {Fanti},
  {Mack}  \& {Govoni}}{{Parma} et~al.}{2007}]{Parmaetal07}
{Parma} P.,  {Murgia} M.,  {de Ruiter} H.~R.,  {Fanti} R.,  {Mack} K.-H.,
  {Govoni} F.,  2007, \mn@doi [A\&A] {10.1051/0004-6361:20077592}, \href
  {http://adsabs.harvard.edu/abs/2007A\%26A...470..875P} {470, 875}

\bibitem[\protect\citeauthoryear{{Perley} \& {Butler}}{{Perley} \&
  {Butler}}{2017}]{Perleyetal17}
{Perley} R.~A.,  {Butler} B.~J.,  2017, \mn@doi [\apjs]
  {10.3847/1538-4365/aa6df9}, \href
  {https://ui.adsabs.harvard.edu/abs/2017ApJS..230....7P} {230, 7}

\bibitem[\protect\citeauthoryear{{Peterson} \& {Fabian}}{{Peterson} \&
  {Fabian}}{2006}]{PetersonFabian06}
{Peterson} J.~R.,  {Fabian} A.~C.,  2006, \mn@doi [Phys. Rep.]
  {10.1016/j.physrep.2005.12.007}, \href
  {http://adsabs.harvard.edu/abs/2006PhR...427....1P} {427, 1}

\bibitem[\protect\citeauthoryear{{Pope}, {Babul}, {Pavlovski}, {Bower}  \&
  {Dotter}}{{Pope} et~al.}{2010}]{Popeetal10}
{Pope} E.~C.~D.,  {Babul} A.,  {Pavlovski} G.,  {Bower} R.~G.,   {Dotter} A.,
  2010, \mn@doi [MNRAS] {10.1111/j.1365-2966.2010.16816.x}, 406, 2023

\bibitem[\protect\citeauthoryear{{Randall}, {Jones}, {Kraft}, {Forman}  \&
  {O'Sullivan}}{{Randall} et~al.}{2009}]{Randalletal09}
{Randall} S.~W.,  {Jones} C.,  {Kraft} R.,  {Forman} W.~R.,   {O'Sullivan} E.,
  2009, \mn@doi [ApJ] {10.1088/0004-637X/696/2/1431}, \href
  {http://adsabs.harvard.edu/abs/2009ApJ...696.1431R} {696, 1431}

\bibitem[\protect\citeauthoryear{{Roediger}, {Br{\"u}ggen}, {Simionescu},
  {B{\"o}hringer}, {Churazov}  \& {Forman}}{{Roediger}
  et~al.}{2011}]{Roedigeretal11}
{Roediger} E.,  {Br{\"u}ggen} M.,  {Simionescu} A.,  {B{\"o}hringer} H.,
  {Churazov} E.,   {Forman} W.~R.,  2011, \mn@doi [MNRAS]
  {10.1111/j.1365-2966.2011.18279.x}, \href
  {http://adsabs.harvard.edu/abs/2011MNRAS.413.2057R} {413, 2057}

\bibitem[\protect\citeauthoryear{{Roediger}, {Kraft}, {Machacek}, {Forman},
  {Nulsen}, {Jones}  \& {Murray}}{{Roediger} et~al.}{2012}]{Roedigeretal12}
{Roediger} E.,  {Kraft} R.~P.,  {Machacek} M.~E.,  {Forman} W.~R.,  {Nulsen}
  P.~E.~J.,  {Jones} C.,   {Murray} S.~S.,  2012, \mn@doi [ApJ]
  {10.1088/0004-637X/754/2/147}, \href
  {http://adsabs.harvard.edu/abs/2012ApJ...754..147R} {754, 147}

\bibitem[\protect\citeauthoryear{{Russell} et~al.,}{{Russell}
  et~al.}{2016}]{Russelletal16}
{Russell} H.~R.,  et~al., 2016, \mn@doi [MNRAS] {10.1093/mnras/stw409}, \href
  {http://adsabs.harvard.edu/abs/2016MNRAS.458.3134R} {458, 3134}

\bibitem[\protect\citeauthoryear{{Russell} et~al.,}{{Russell}
  et~al.}{2019}]{Russelletal19}
{Russell} H.~R.,  et~al., 2019, MNRAS, 490, 3025

\bibitem[\protect\citeauthoryear{{Salom{\'e}}, {Combes}, {Revaz}, {Downes},
  {Edge}  \& {Fabian}}{{Salom{\'e}} et~al.}{2011}]{Salomeetal11}
{Salom{\'e}} P.,  {Combes} F.,  {Revaz} Y.,  {Downes} D.,  {Edge} A.~C.,
  {Fabian} A.~C.,  2011, \mn@doi [A\&A] {10.1051/0004-6361/200811333}, \href
  {http://adsabs.harvard.edu/abs/2011A\%26A...531A..85S} {531, A85}

\bibitem[\protect\citeauthoryear{{Sanders}, {Fabian}, {Allen}  \&
  {Schmidt}}{{Sanders} et~al.}{2004}]{Sandersetal04}
{Sanders} J.~S.,  {Fabian} A.~C.,  {Allen} S.~W.,   {Schmidt} R.~W.,  2004,
  \mn@doi [MNRAS] {10.1111/j.1365-2966.2004.07576.x}, \href
  {http://adsabs.harvard.edu/cgi-bin/nph-bib_query?bibcode=2004MNRAS.349..952S&db_key=AST}
  {349, 952}

\bibitem[\protect\citeauthoryear{{Sanderson}, {Ponman}  \&
  {O'Sullivan}}{{Sanderson} et~al.}{2006}]{Sandersonetal06}
{Sanderson} A.~J.~R.,  {Ponman} T.~J.,   {O'Sullivan} E.,  2006, \mn@doi
  [MNRAS] {10.1111/j.1365-2966.2006.10956.x}, \href
  {http://adsabs.harvard.edu/cgi-bin/nph-bib_query?bibcode=2006MNRAS.372.1496S&db_key=AST}
  {372, 1496}

\bibitem[\protect\citeauthoryear{{Scaife} \& {Heald}}{{Scaife} \&
  {Heald}}{2012}]{ScaifeHeald12}
{Scaife} A.~M.~M.,  {Heald} G.~H.,  2012, \mn@doi [\mnras]
  {10.1111/j.1745-3933.2012.01251.x}, \href
  {http://cdsads.u-strasbg.fr/abs/2012MNRAS.423L..30S} {423, L30}

\bibitem[\protect\citeauthoryear{{Schellenberger}, {Vrtilek}, {David},
  {O'Sullivan}, {Giacintucci}, {Johnston-Hollitt}, {Duchesne}  \&
  {Raychaudhury}}{{Schellenberger} et~al.}{2017}]{Schellenbergeretal17}
{Schellenberger} G.,  {Vrtilek} J.~M.,  {David} L.,  {O'Sullivan} E.,
  {Giacintucci} S.,  {Johnston-Hollitt} M.,  {Duchesne} S.~W.,   {Raychaudhury}
  S.,  2017, \mn@doi [ApJ] {10.3847/1538-4357/aa7f2e}, 845, 84

\bibitem[\protect\citeauthoryear{{Simionescu}, {Werner}, {B{\"o}hringer},
  {Kaastra}, {Finoguenov}, {Br{\"u}ggen}  \& {Nulsen}}{{Simionescu}
  et~al.}{2009}]{Simionescuetal09}
{Simionescu} A.,  {Werner} N.,  {B{\"o}hringer} H.,  {Kaastra} J.~S.,
  {Finoguenov} A.,  {Br{\"u}ggen} M.,   {Nulsen} P.~E.~J.,  2009, \mn@doi
  [A\&A] {10.1051/0004-6361:200810225}, \href
  {http://adsabs.harvard.edu/abs/2009A\%26A...493..409S} {493, 409}

\bibitem[\protect\citeauthoryear{{Su} et~al.,}{{Su} et~al.}{2017}]{Suetal17}
{Su} Y.,  et~al., 2017, \mn@doi [ApJ] {10.3847/1538-4357/aa989e}, \href
  {http://adsabs.harvard.edu/abs/2017ApJ...851...69S} {851, 69}

\bibitem[\protect\citeauthoryear{{Sun}, {Forman}, {Vikhlinin}, {Hornstrup},
  {Jones}  \& {Murray}}{{Sun} et~al.}{2003}]{Sunetal03}
{Sun} M.,  {Forman} W.,  {Vikhlinin} A.,  {Hornstrup} A.,  {Jones} C.,
  {Murray} S.~S.,  2003, \mn@doi [ApJ] {10.1086/378887}, \href
  {http://adsabs.harvard.edu/cgi-bin/nph-bib_query?bibcode=2003ApJ...598..250S&db_key=AST}
  {598, 250}

\bibitem[\protect\citeauthoryear{{Sun}, {Voit}, {Donahue}, {Jones}, {Forman}
  \& {Vikhlinin}}{{Sun} et~al.}{2009}]{Sunetal09}
{Sun} M.,  {Voit} G.~M.,  {Donahue} M.,  {Jones} C.,  {Forman} W.,
  {Vikhlinin} A.,  2009, \mn@doi [ApJ] {10.1088/0004-637X/693/2/1142}, \href
  {http://adsabs.harvard.edu/abs/2009ApJ...693.1142S} {693, 1142}

\bibitem[\protect\citeauthoryear{Toomre \& Toomre}{Toomre \&
  Toomre}{1972}]{Toomres72}
Toomre A.,  Toomre J.,  1972, ApJ, 178, 623

\bibitem[\protect\citeauthoryear{{Tremblay} et~al.,}{{Tremblay}
  et~al.}{2018}]{Tremblayetal18}
{Tremblay} G.~R.,  et~al., 2018, \mn@doi [ApJ] {10.3847/1538-4357/aad6dd},
  \href {http://adsabs.harvard.edu/abs/2018ApJ...865...13T} {865, 13}

\bibitem[\protect\citeauthoryear{{Venturi}, {Rossetti}, {Bardelli},
  {Giacintucci}, {Dallacasa}, {Cornacchia}  \& {Kantharia}}{{Venturi}
  et~al.}{2013}]{Venturietal13}
{Venturi} T.,  {Rossetti} M.,  {Bardelli} S.,  {Giacintucci} S.,  {Dallacasa}
  D.,  {Cornacchia} M.,   {Kantharia} N.~G.,  2013, \mn@doi [\aap]
  {10.1051/0004-6361/201322023}, \href
  {https://ui.adsabs.harvard.edu/abs/2013A&A...558A.146V} {558, A146}

\bibitem[\protect\citeauthoryear{{Weisskopf}, {Brinkman}, {Canizares},
  {Garmire}, {Murray}  \& {Van Speybroeck}}{{Weisskopf}
  et~al.}{2002}]{Weisskopfetal02}
{Weisskopf} M.~C.,  {Brinkman} B.,  {Canizares} C.,  {Garmire} G.,  {Murray}
  S.,   {Van Speybroeck} L.~P.,  2002, PASP, \href
  {http://adsabs.harvard.edu/cgi-bin/nph-bib_query?bibcode=2002PASP..114....1W&db_key=AST}
  {114, 1}

\bibitem[\protect\citeauthoryear{{Worrall} \& {Birkinshaw}}{{Worrall} \&
  {Birkinshaw}}{2006}]{WorrallBirkinshaw06}
{Worrall} D.~M.,  {Birkinshaw} M.,  2006, in {D.~Alloin} ed.,  Lecture Notes in
  Physics, Berlin Springer Verlag Vol. 693, Physics of Active Galactic Nuclei
  at all Scales. p.~39 (\mn@eprint {} {arXiv:astro-ph/0410297})

\bibitem[\protect\citeauthoryear{{Zhang} et~al.,}{{Zhang}
  et~al.}{2017}]{Zhangetal17}
{Zhang} Y.-Y.,  et~al., 2017, A\&A, \href
  {https://ui-adsabs-harvard-edu.ezp-prod1.hul.harvard.edu/abs/2017A&A...599A.138Z}
  {599, A138}

\bibitem[\protect\citeauthoryear{{ZuHone}, {Markevitch}  \& {Lee}}{{ZuHone}
  et~al.}{2011}]{ZuHoneetal11}
{ZuHone} J.~A.,  {Markevitch} M.,   {Lee} D.,  2011, \mn@doi [ApJ]
  {10.1088/0004-637X/743/1/16}, \href
  {http://adsabs.harvard.edu/abs/2011ApJ...743...16Z} {743, 16}

\bibitem[\protect\citeauthoryear{{ZuHone}, {Kowalik}, {{\"O}hman}, {Lau}  \&
  {Nagai}}{{ZuHone} et~al.}{2018}]{ZuHoneetal18}
{ZuHone} J.~A.,  {Kowalik} K.,  {{\"O}hman} E.,  {Lau} E.,   {Nagai} D.,  2018,
  \mn@doi [ApJS] {10.3847/1538-4365/aa99db}, \href
  {http://adsabs.harvard.edu/abs/2018ApJS..234....4Z} {234, 4}

\makeatother
\end{thebibliography}



\bsp	
\label{lastpage}
\end{document}